\documentclass[aps,prc,twocolumn,showpacs,superscriptaddress,amsmath,amssymb,showkeys]{revtex4}
\usepackage{graphicx}% Include figure files
\usepackage{dcolumn}% Align table columns on decimal point
\usepackage{bm}% bold math
\begin{document}
%\preprint{APS/123-QED}
%
\title{Generalized Dalitz Plot analysis of the near threshold $pp\rightarrow ppK^{+}K^{-}$ reaction in view of the $K^{+}K^{-}$ final state interaction}
\author{M.~Silarski}
\affiliation{Institute of Physics, Jagiellonian University, PL-30-059 Cracow, Poland}
\author{P.~Moskal} \email[Electronic address: ]{p.moskal@fz-juelich.de}
\affiliation{Institute of Physics, Jagiellonian University, PL-30-059 Cracow, Poland}
\affiliation{Institute for Nuclear Physics and J{\"u}lich Center for Hadron Physics,\\
                 Research Center J{\"u}lich, D-52425 J{\"u}lich, Germany}
\author{E.~Czerwi\'nski}
\affiliation{Institute of Physics, Jagiellonian University, PL-30-059 Cracow, Poland}
\affiliation{Institute for Nuclear Physics and J{\"u}lich Center for Hadron Physics,\\
                 Research Center J{\"u}lich, D-52425 J{\"u}lich, Germany}
\author{R.~Czy\.zykiewicz}
\affiliation{Institute of Physics, Jagiellonian University, PL-30-059 Cracow, Poland}
\author{D.~Gil}
\affiliation{Institute of Physics, Jagiellonian University, PL-30-059 Cracow, Poland}
\author{D.~Grzonka}
\affiliation{Institute for Nuclear Physics and J{\"u}lich Center for Hadron Physics,\\
                 Research Center J{\"u}lich, D-52425 J{\"u}lich, Germany}
\author{L.~Jarczyk}
\affiliation{Institute of Physics, Jagiellonian University, PL-30-059 Cracow, Poland}
\author{B.~Kamys}
\affiliation{Institute of Physics, Jagiellonian University, PL-30-059 Cracow, Poland}
\author{A.~Khoukaz}
\affiliation{IKP, Westf\"alische Wilhelms-Universit\"at, D-48149 M\"unster, Germany}
\author{J.~Klaja}
\affiliation{Institute of Physics, Jagiellonian University, PL-30-059 Cracow, Poland}
\affiliation{Institute for Nuclear Physics and J{\"u}lich Center for Hadron Physics,\\
                 Research Center J{\"u}lich, D-52425 J{\"u}lich, Germany}
\author{P.~Klaja}
\affiliation{Institute of Physics, Jagiellonian University, PL-30-059 Cracow, Poland}
\affiliation{Institute for Nuclear Physics and J{\"u}lich Center for Hadron Physics,\\
                 Research Center J{\"u}lich, D-52425 J{\"u}lich, Germany}
\author{W.~Krzemie{\'n}}
\affiliation{Institute of Physics, Jagiellonian University, PL-30-059 Cracow, Poland}
\affiliation{Institute for Nuclear Physics and J{\"u}lich Center for Hadron Physics,\\
                 Research Center J{\"u}lich, D-52425 J{\"u}lich, Germany}
\author{W.~Oelert}
\affiliation{Institute for Nuclear Physics and J{\"u}lich Center for Hadron Physics,\\
                 Research Center J{\"u}lich, D-52425 J{\"u}lich, Germany}
\author{J.~Ritman}
\affiliation{Institute for Nuclear Physics and J{\"u}lich Center for Hadron Physics,\\
                 Research Center J{\"u}lich, D-52425 J{\"u}lich, Germany}
\author{T.~Sefzick}
\affiliation{Institute for Nuclear Physics and J{\"u}lich Center for Hadron Physics,\\
                 Research Center J{\"u}lich, D-52425 J{\"u}lich, Germany}
\author{M.~Siemaszko}
\affiliation{Institute of Physics, University of Silesia, PL-40-007 Katowice, Poland}
\author{J.~Smyrski}
\affiliation{Institute of Physics, Jagiellonian University, PL-30-059 Cracow, Poland}
\author{A.~T\"aschner}
\affiliation{IKP, Westf\"alische Wilhelms-Universit\"at, D-48149 M\"unster, Germany}
\author{P.~Winter}
\affiliation{University of Illinois at Urbana-Champaign, Urbana, IL 61801, USA}
\author{M.~Wolke}
\affiliation{Institute for Nuclear Physics and J{\"u}lich Center for Hadron Physics,\\
                 Research Center J{\"u}lich, D-52425 J{\"u}lich, Germany}
\author{P.~W\"ustner}
\affiliation{Institute for Nuclear Physics and J{\"u}lich Center for Hadron Physics,\\
                 Research Center J{\"u}lich, D-52425 J{\"u}lich, Germany}
\author{M.~Zieli\'nski}
\affiliation{Institute of Physics, Jagiellonian University, PL-30-059 Cracow, Poland}
\author{W.~Zipper}
\affiliation{Institute of Physics, University of Silesia, PL-40-007 Katowice, Poland}
\author{J.~Zdebik}
\affiliation{Institute of Physics, Jagiellonian University, PL-30-059 Cracow, Poland}
\date{\today}% It is always \today, today,
\begin{abstract}

The excitation function for the $pp\to ppK^+K^-$ reaction revealed a
significant enhancement close to threshold which may plausibly be assigned
to the influence of the $pK^-$ and $K^+K^-$ final state interactions.
In an improved reanalysis of COSY-11 data for the $pp\to ppK^+K^-$ reaction
at excess energies of Q~=~10 MeV and 28 MeV including the proton-$K^-$
interaction the enhancement is confirmed. Invariant mass distributions for
the two- and three-particle subsystems allow to test at low excess energies
the ansatz and parameters for the description of the interaction in the $ppK^+K^-$
system as derived from the COSY-ANKE data.
Finally, based for the first time on the low energy $K^{+}K^{-}$ invariant
mass distributions and the generalized Dalitz plot analysis, we estimate
the scattering length for the  $K^{+}K^{-}$ interaction
to be $|Re({a_{K^+K^-})}|$ = 0.5$^{+4.0}_{-0.5}$~fm
and $Im({a_{K^+K^-}})$ = 3.0 $\pm$ 3.0~fm.

\vspace{1pc}
\end{abstract}
\maketitle
\section{Introduction}
The basic motivation for investigating the $pp\to ppK^+K^-$ reaction near the kinematical 
threshold is comprehensively reviewed in~\cite{oelert}, as an attempt to understand the nature
of the scalar resonances
$f_{0}$(980) and $a_{0}$(980), whose masses are very close to the
sum of the $K^{+}$ and $K^{-}$ masses. Besides the standard
interpretation as $q\bar{q}$ mesons~\cite{Morgan}, these
resonances were also proposed to be $qq\bar{q}\bar{q}$
states~\cite{Jaffe}, $K\bar{K}$ molecules~\cite{Lohse, Weinstein},
hybrid $q\bar{q}$/meson-meson systems~\cite{Beveren} or even quark-less
gluonic hadrons ~\cite{Johnson}.
The strength of the $K\bar{K}$ interaction is a crucial quantity regarding the formation
of a $K\bar{K}$ molecule, whereas the $KN$ interaction is of importance in view of 
the vigorous discussion concerning the structure of the excited hyperon $\Lambda$(1405) 
which is considered as a three quark system or as a $KN$ molecular state~\cite{Kaiser}.
Additionally, these interactions appear to be very important also with respect to other
phenomena, like possible kaon condensation in neutron stars~\cite{Li}, or production of strange particles
immersed in a dense nuclear medium studied by means of heavy ion collisions~\cite{Senger,Laue,Barth,Menzel}.

Measurements of the near threshold $pp\to ppK^+K^-$ reaction have been 
made possible by beams of low emittance and small
momentum spread available at storage ring facilities, in
particular at the cooler synchrotron COSY at the research center in J{\"u}lich in Germany~\cite{cosy}.
A precise determination of the collision energy, in the order of  fractions of MeV,
permitted to deal with  the rapid growth of cross sections~\cite{review} and thus to
take advantage of the threshold kinematics like, e.g., \ full phase space coverage
achievable with dipole magnetic spectrometers being rather limited in geometrical acceptance.
Early experiments on $K^{+}K^{-}$ pair production at COSY conducted by the COSY-11
collaboration revealed, however,  that the total cross section at threshold is by more
than seven orders of magnitude smaller than the total proton-proton production
cross section making the study difficult due to low statistics~\cite{wolke,quentmeier,winter}.
A possible influence from the $f_{0}$ or $a_{0}$ mesons on the $K^{+}K^{-}$ pair production
appeared to be too weak to be distinguished from the direct production of these mesons 
based on the COSY-11 data~\cite{quentmeier}. Recent results obtained by the ANKE collaboration
with much higher statistics can also be explained without the need of including the scalars
$f_{0}$ or $a_{0}$~\cite{anke,c_wilkin}.
However, the combined systematic collection of  data below~\cite{wolke,quentmeier,winter}
and above~\cite{anke,disto} the $\phi$ meson threshold reveal a significant signal
in the shape of the excitation function
which  may be due to the $K^{-}p$ and perhaps also to the $K^{+}K^{-}$ interaction.
This signal is based on the COSY-11 data which, as indicated by
authors of article~\cite{anke}, were analyzed calculating the acceptance without
the inclusion of the $pK^-$ interaction. Therefore a more detailed analysis of the COSY-11 data
at excess energies of Q~=~10~MeV and 28~MeV including now studies of both the differential cross
section distributions and the strength of the final state interaction between the $K^{+}$ and
$K^{-}$ mesons was performed.
The analysis is based on a generalization of the Dalitz plot for four particles as proposed by
Goldhaber \textit{et al.}~\cite{goldhaber1,goldhaber2,wilkin2007}.
%#######
\section{Excitation function for the near threshold $pp\to ppK^{+}K^{-}$ reaction}
\begin{figure}[h]
\centering 
\includegraphics[width=0.48\textwidth,angle=0]{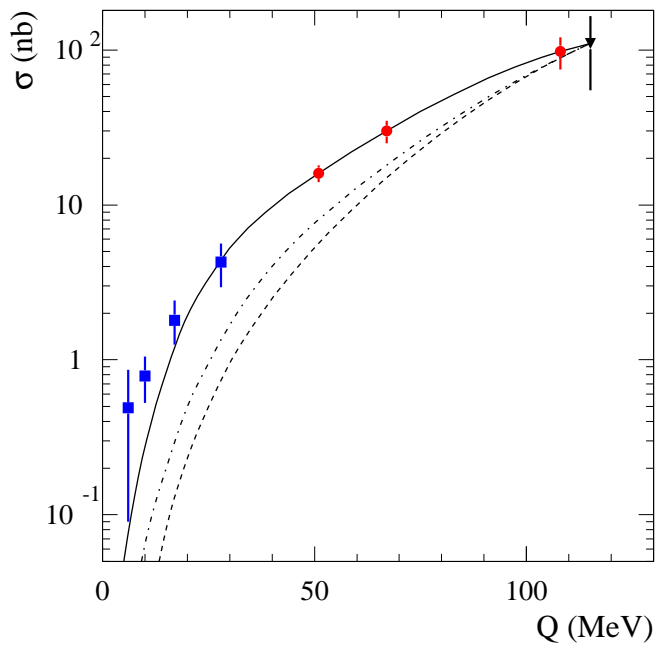}
\caption{(Color online) Excitation function for the $pp\rightarrow ppK^{+}K^{-}$ reaction.
Triangle and circles represent the DISTO and ANKE measurements, respectively.
The four points close to the threshold are results from the COSY-11 measurements.
The curves are described in the text.}
\label{excitation-f}
\end{figure}

%#######
The measurements of the $pp\rightarrow ppK^{+}K^{-}$ reaction were
conducted at low excess energies by the collaborations ANKE~\cite{anke},
COSY-11~\cite{quentmeier,winter,wolke} and DISTO~\cite{disto}.
The achieved results are presented in Fig.~\ref{excitation-f} 
together with curves representing three different theoretical expectations~\cite{anke} 
normalized to the DISTO data point at Q~=~114~MeV.
The dashed curve represents the energy dependence from four-body phase space when we assume that
there is no interaction between particles in the final state. These calculations differ
from the experimental data by two orders of magnitude at Q~=~10~MeV and by a factor of
about five at Q~=~28~MeV. Hence, it is obvious, that effects of final state interactions
cannot be neglected in the $ppK^{+}K^{-}$ system~\cite{oelert1}.
Inclusion of the $pp$--FSI (dashed-dotted line in Fig.~\ref{excitation-f}), by folding
its parameterization known from the three body final state~\cite{pp-FSI} with the four
body phase space, is closer to the experimental data, but does not fully
account for the difference~\cite{winter}. The enhancement may be due to the influence of
$pK$ and $K^{+}K^{-}$ interaction which was neglected in the calculations. Indeed, as shown
by authors of reference~\cite{anke,c_wilkin}
the inclusion of the $pK^{-}$--FSI (solid line) reproduces the experimental data for
excess energies down to Q~=~28~MeV. These calculations of the cross section were accomplished
under the assumption that the overall enhancement factor, originating from final state
interaction in the $ppK^{+}K^{-}$ system, can be factorised into enhancements in the
$pp$ and two $pK^{-}$ subsystems~\cite{anke}:
 \begin{equation}
F_{FSI}~=~F_{pp}(q)\cdot F_{p_{1}K^{-}}(k_{1})\cdot F_{p_{2}K^{-}}(k_{2})~,
\label{pp-pkfsi}
\end{equation}
where $k_1$, $k_2$ and $q$ stand for the relative momenta of the particles in the first $pK^-$
subsystem, second $pK^-$ subsystem and $pp$ subsystem, respectively. The factors describing
the enhancement originating from the $pK^{-}$--FSI are parametrized using the scattering
length approximation.
It is important to note that the inclusion of the $pp$ and $pK^{-}$ final state interaction 
is not sufficient to describe the data very close to threshold (see Fig.~\ref{excitation-f}).
This enhancement may be due to the influence of the $K^{+}K^{-}$
interaction, which was neglected in the calculations~\footnote
{
It is worth mentioning, that in the calculations also the $pK^{+}$ 
interaction was neglected. It is repulsive and weak and hence it can be 
interpreted as an additional attraction in the $pK^{-}$ system~\cite{anke}.
}.
However, as pointed out in Ref.~\cite{anke} the observed increase of the total
cross section near threshold may be due to the neglect of the
$pK^{-}$--FSI in the calculations of the COSY-11 acceptance.
As a consequence the extracted cross sections would decrease, if this interaction was
taken into account during the analysis of the experimental data.
This concern encouraged us to check quantitatively the influence
of the interaction in the $pK^{-}$ subsystem on the acceptance of
the detection setup.
In addition, absolute values for the differential distributions
of the $pK$ and $ppK$ invariant masses were extracted and generalized Dalitz plot analysis
of the data in view of the $K^{+}K^{-}$ interaction, was performed.
%######
\section{Measurements of the $pp\to ppK^{+}K^{-}$ reaction performed with the COSY-11 magnetic spectrometer}
\begin{figure}[h]
\includegraphics[angle=270,width=0.51\textwidth]{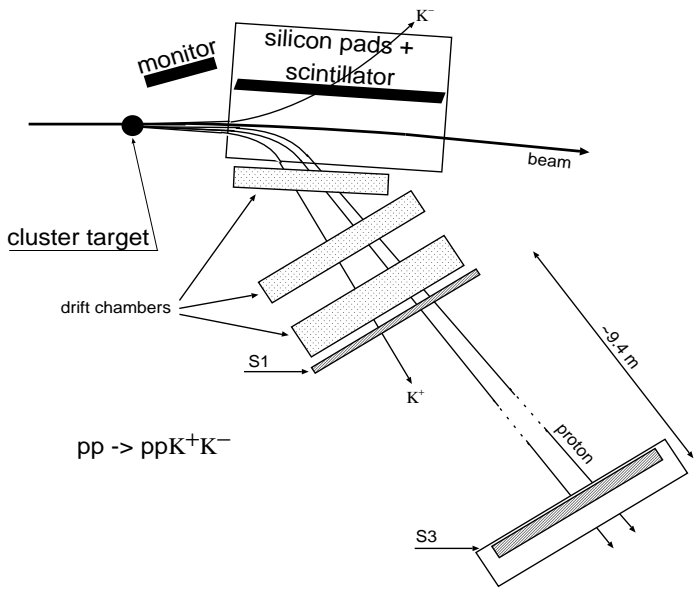}
\caption{Schematic view of the COSY-11 detector with a typical event of
the $pp\rightarrow ppK^{+}K^{-}$ reaction channel. For the description
see text.}
\label{detector}
\end{figure}

The measurements of the $pp\rightarrow ppK^{+}K^{-}$  reaction close to threshold
have been conducted using the cooler synchrotron COSY~\cite{cosy} and the
COSY-11 detector system~\cite{c-11} shown schematically in Fig.~\ref{detector}.
The target, being a beam of H$_2$ molecules grouped to clusters of up to 10$^5$ atoms~\cite{dombrowski},
crosses perpendicularly the proton beam circulating in the ring.
If a collision of protons leads to the production of a $K^{+}K^{-}$ meson pair,
then the reaction products, having smaller momenta than the circulating beam,
are directed by the magnetic dipole field towards the COSY-11 detection system
and leave the vacuum chamber through a thin exit foil~\cite{c-11}.
Tracks of positively charged particles, registered by  drift chambers, are traced back
through the magnetic field to the nominal interaction point leading to a momentum
determination. The knowledge of the momentum combined with
an independent measurement of the velocity, performed by means of the
scintillation detectors S1 and S3, permits to identify the registered particles and
to determine their four momentum vectors.
Knowing both the four momenta of the three positively charged ejectiles and the
proton beam momentum one can calculate the mass of the unobserved system $X^{-}$.
In the case of the $pp\rightarrow ppK^{+}K^{-}$ reaction this should correspond to
the mass of the $K^{-}$ meson, but we observe also background originating partly from the
$pp\rightarrow pp\pi^{+}X^{-}$ reaction, where the $\pi^{+}$ was misidentified
as a $K^{+}$ meson, and in part due to the $K^{+}$ meson being associated with the
hyperons $\Lambda$(1405) or $\Sigma$(1385) production~\cite{quentmeier,winter}.
This background, however, can be completely removed by demanding a signal
in the silicon pad detectors (mounted inside the dipole) at the position where
the $K^{-}$ meson originating from the $pp\rightarrow ppK^{+}K^{-}$ reaction
is expected. This clear identification allows to select the
$pp\rightarrow ppK^{+}K^{-}$ events and to determine the total and differential
cross sections. A more detailed description of the experiment and data
evaluation is given in~\cite{c-11,winter,
quentmeier,moskal1,wolke}.
%#######
\section{Differential observables for COSY-11 data at Q~=~10~M\lowercase{e}V and Q = 28 M\lowercase{e}V}
%#############
\begin{table}
\caption{The differential cross sections of the $pp\to ppK^+K^-$ reaction
as a function of invariant masses for different subsystems.
The values of $M_{ij}$ denote the center of the invariant mass bins of 2.5~MeV/$\text{c}^2$ and
7.0~MeV/$\text{c}^2$ width for Q~=~10~MeV and Q~=~28~MeV, respectively.}
\label{tab_all}
\begin{ruledtabular}
\begin{tabular}{ccc|cc}
&\multicolumn{2}{c|}{Q~=~10~MeV}&\multicolumn{2}{c}{Q~=~28~MeV}\\
\hline
&$M_{pp}$~($\frac{\text{GeV}}{\text{c}^2}$) &$\frac{d\sigma}{dM_{pp}}$~($\frac{\text{nb}}{\text{GeV/c}^2}$)
&$M_{pp}$~($\frac{\text{GeV}}{\text{c}^2}$) &$\frac{d\sigma}{dM_{pp}}$~($\frac{\text{nb}}{\text{GeV/c}^2}$)\\
\hline
&1.8778  &216~$\pm$~53 &1.880 &388~$\pm$~92\\
&1.8803  &106~$\pm$~34 &1.887 &346~$\pm$~87\\
&1.8828  &43~$\pm$~22 &1.894 &148~$\pm$~67\\
&1.8853  &14~$\pm$~14 &1.901 &48~$\pm$~48\\
\hline
&$M_{pK^+}$~($\frac{\text{GeV}}{\text{c}^2}$) &$\frac{d\sigma}{dM_{pK^+}}$~($\frac{\text{nb}}{\text{GeV/c}^2}$)
&$M_{pK^+}$~($\frac{\text{GeV}}{\text{c}^2}$) &$\frac{d\sigma}{dM_{pK^+}}$~($\frac{\text{nb}}{\text{GeV/c}^2}$)\\
\hline
&1.4332 &107~$\pm$~34 &1.435 &255~$\pm$~72\\
&1.4357 &122~$\pm$~39 &1.442 &261~$\pm$~76\\
&1.4382 &125~$\pm$~42 &1.449 &287~$\pm$~83\\
&1.4407 &32~$\pm$~21 &1.456 &63~$\pm$~37\\
\hline
&$M_{pK^-}$~($\frac{\text{GeV}}{\text{c}^2}$) &$\frac{d\sigma}{dM_{pK^-}}$~($\frac{\text{nb}}{\text{GeV/c}^2}$)
&$M_{pK^-}$~($\frac{\text{GeV}}{\text{c}^2}$) &$\frac{d\sigma}{dM_{pK^-}}$~($\frac{\text{nb}}{\text{GeV/c}^2}$)\\
\hline
&1.4332 &173$~\pm$~47 &1.435 &581~$\pm$~117\\
&1.4357 &145~$\pm$~42 &1.442 &135~$\pm$~56\\
&1.4382 &36~$\pm$~21 &1.449 &97~$\pm$~46\\
&1.4407 &13~$\pm$~11 &1.456 &25~$\pm$~21\\
\hline
&$M_{K^+K^-}$~($\frac{\text{GeV}}{\text{c}^2}$) &$\frac{d\sigma}{dM_{K^+K^-}}$~($\frac{\text{nb}}{\text{GeV/c}^2}$)
&$M_{K^+K^-}$~($\frac{\text{GeV}}{\text{c}^2}$) &$\frac{d\sigma}{dM_{K^+K^-}}$~($\frac{\text{nb}}{\text{GeV/c}^2}$)\\
\hline
&0.9887 &169~$\pm$~44 &0.991 &221~$\pm$~70\\
&0.9912 &174~$\pm$~51 &0.998 &454~$\pm$~114\\
&0.9937 &35~$\pm$~21 &1.005 &230~$\pm$~70\\
&0.9962 &0~$\pm$~9 &1.012 &38~$\pm$~22\\
\hline
&$M_{ppK^+}$~($\frac{\text{GeV}}{\text{c}^2}$) &$\frac{d\sigma}{dM_{ppK^+}}$~($\frac{\text{nb}}{\text{GeV/c}^2}$)
&$M_{ppK^+}$~($\frac{\text{GeV}}{\text{c}^2}$) &$\frac{d\sigma}{dM_{ppK^+}}$~($\frac{\text{nb}}{\text{GeV/c}^2}$)\\
\hline
&2.3715 &0~$\pm$~13 &2.374 &20~$\pm$~20\\
&2.3740 &68~$\pm$~28 &2.381 &61~$\pm$~36\\
&2.3765 &164~$\pm$~43 &2.388 &247~$\pm$~72\\
&2.3790 &99~$\pm$~38 &2.395 &566~$\pm$~121\\
\hline
&$M_{ppK^-}$~($\frac{\text{GeV}}{\text{c}^2}$) &$\frac{d\sigma}{dM_{ppK^-}}$~($\frac{\text{nb}}{\text{GeV/c}^2}$)
&$M_{ppK^-}$~($\frac{\text{GeV}}{\text{c}^2}$) &$\frac{d\sigma}{dM_{ppK^-}}$~($\frac{\text{nb}}{\text{GeV/c}^2}$)\\
\hline
&2.3715 &60~$\pm$~30 &2.374 &204~$\pm$~68\\
&2.3740 &115~$\pm$~39 &2.381 &284~$\pm$~79\\
&2.3765 &127~$\pm$~39 &2.388 &176~$\pm$~63\\
&2.3790 &87~$\pm$~31 &2.395 &216~$\pm$~69\\
\end{tabular}
\end{ruledtabular}
\end{table}
%$$$$$$$$
In order to check the sensitivity of the result to the assumption of the $pK^{-}$ final state interaction
we derived the distributions of the differential cross section 
assuming that the acceptance depends only on the $pp$--FSI~\footnote
{In all calculations we used the following parametrization of the proton-proton scattering amplitude:
\[F_{pp} =
  \frac{e^{-i\delta_{pp}({^{1}\mbox{\scriptsize S}_{0}})} \cdot
        \sin{\delta_{pp}({^{1}\mbox{S}_0})}}
       {C \cdot \mbox{q}}~,\]
where $C$ stands for the square root of the Coulomb penetration factor~\cite{pp-FSI}.
The parameter $\delta_{pp}({^{1}\mbox{S}_0})$ denotes the phase-shift calculated according
 to the modified Cini-Fubini-Stanghellini formula with the Wong-Noyes Coulomb
correction~\cite{noyes995,naisse506,noyes465}. A more detailed description of this parametrization
can be found in references~\cite{pp-FSI,habilitacja,noyes995,naisse506,noyes465}.}.
Then we calculated the acceptance with inclusion of the $pp$-- and $pK^{-}$--FSI,
and derived analogous distributions. In this calculations we assumed the factorisation of the
final state interaction given by Eq.~\ref{pp-pkfsi} and used the $pK^{-}$ scattering length
$a_{pK^{-}} = (0 + 1.5i)$ fm~\cite{anke}.
\begin{figure}
\centering 
\includegraphics[width=0.23\textwidth,angle=0]{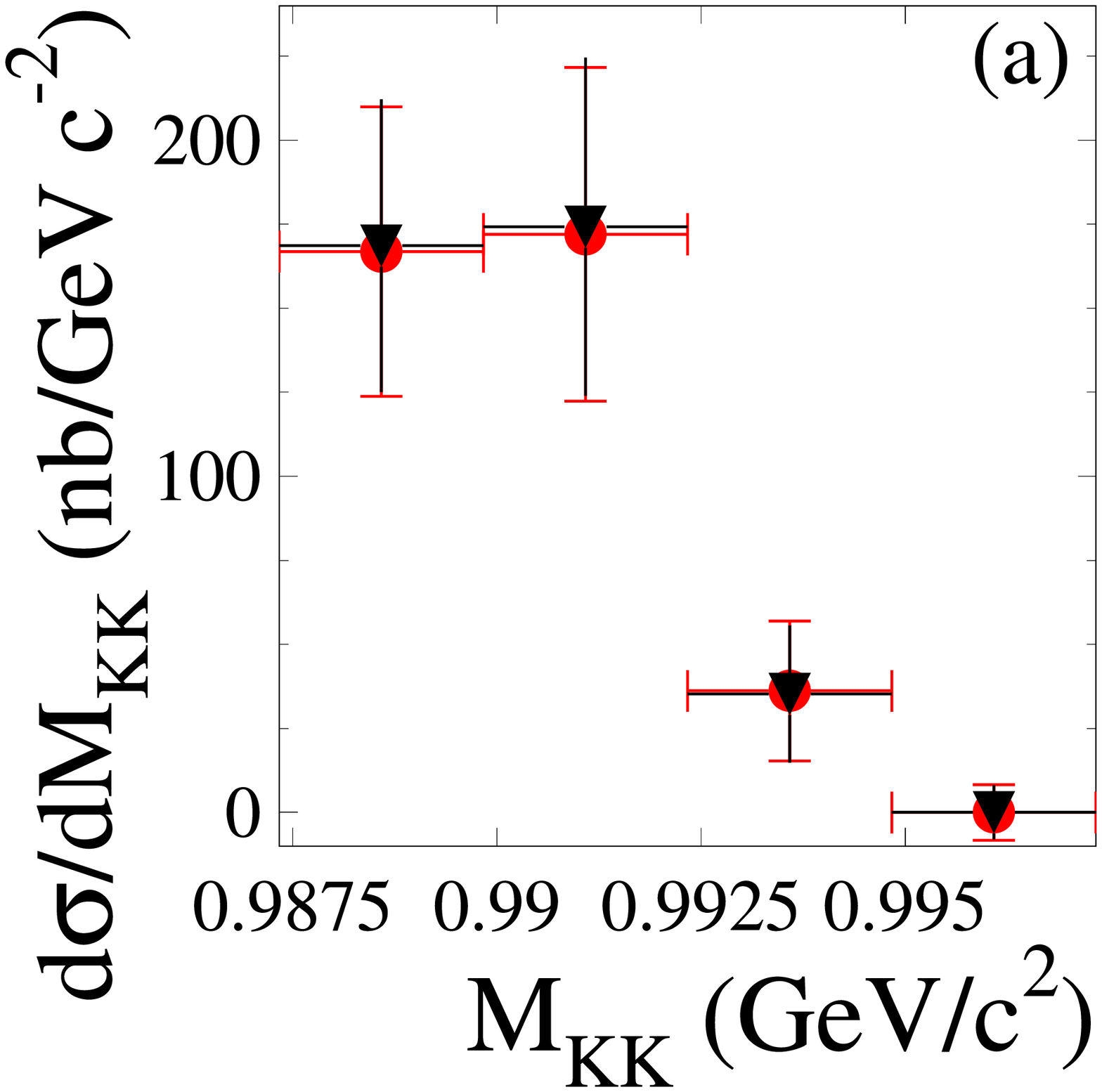}
\includegraphics[width=0.23\textwidth,angle=0]{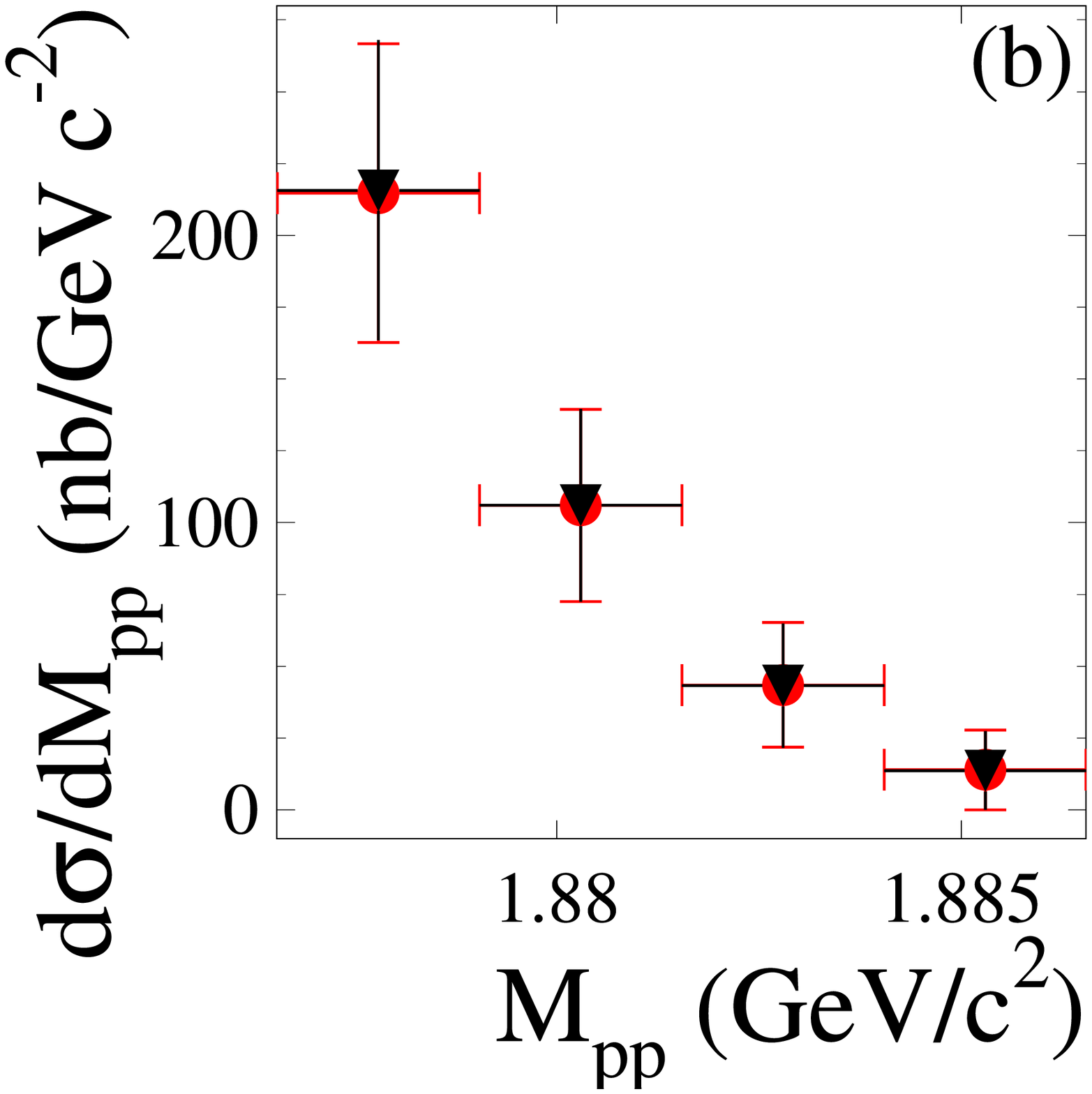}
\includegraphics[width=0.23\textwidth,angle=0]{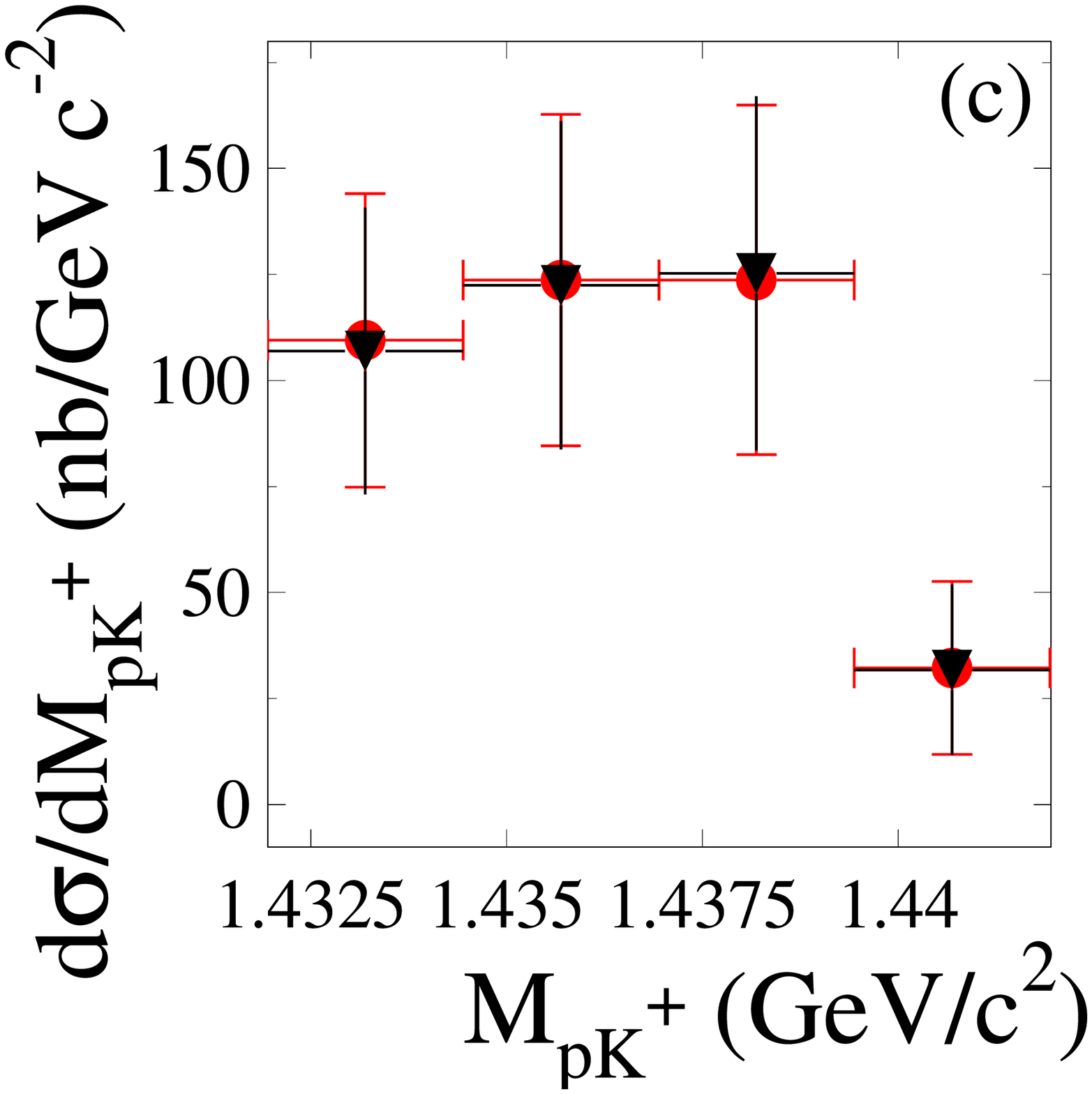}
\includegraphics[width=0.23\textwidth,angle=0]{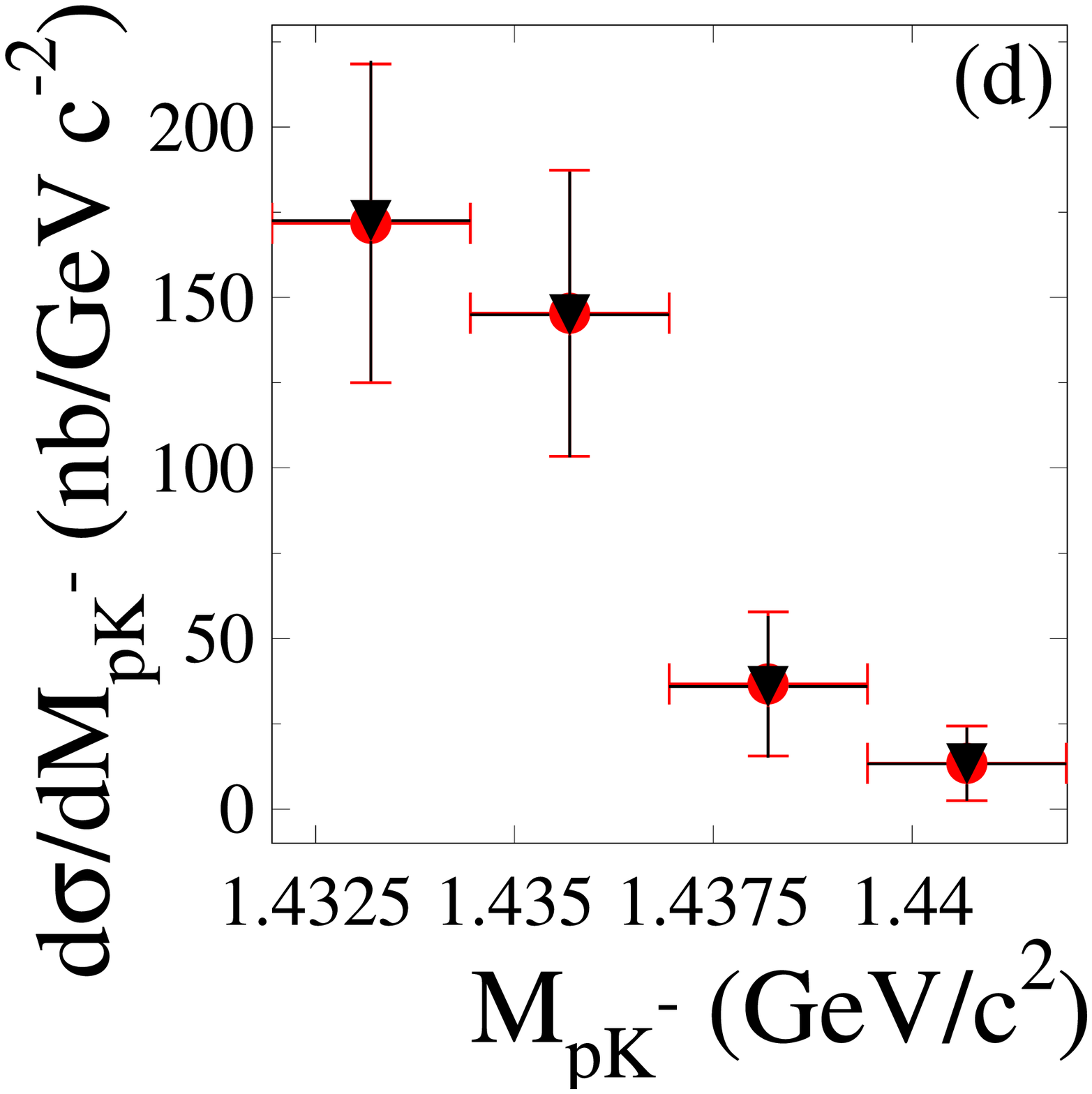}
\includegraphics[width=0.23\textwidth,angle=0]{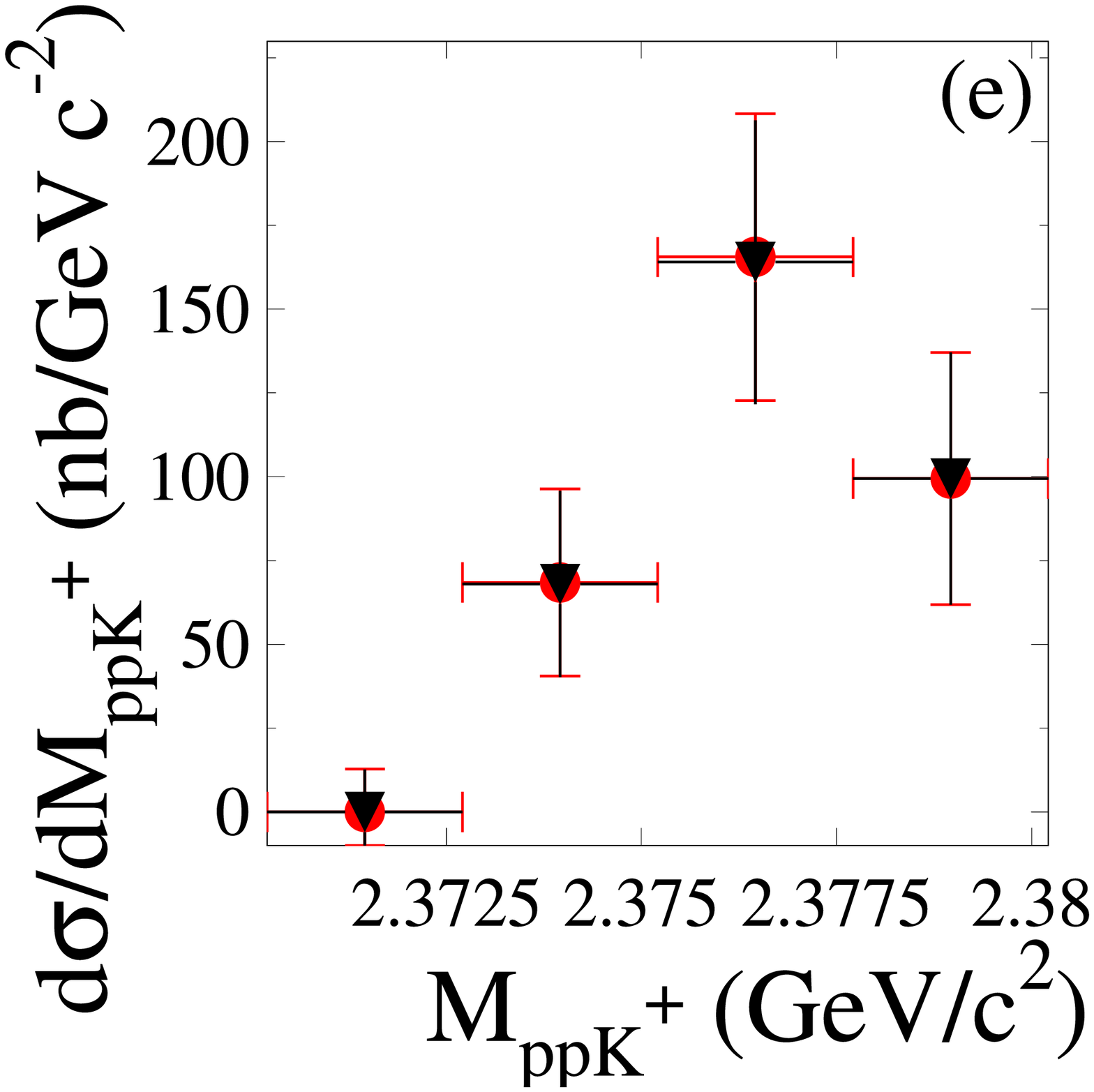}
\includegraphics[width=0.23\textwidth,angle=0]{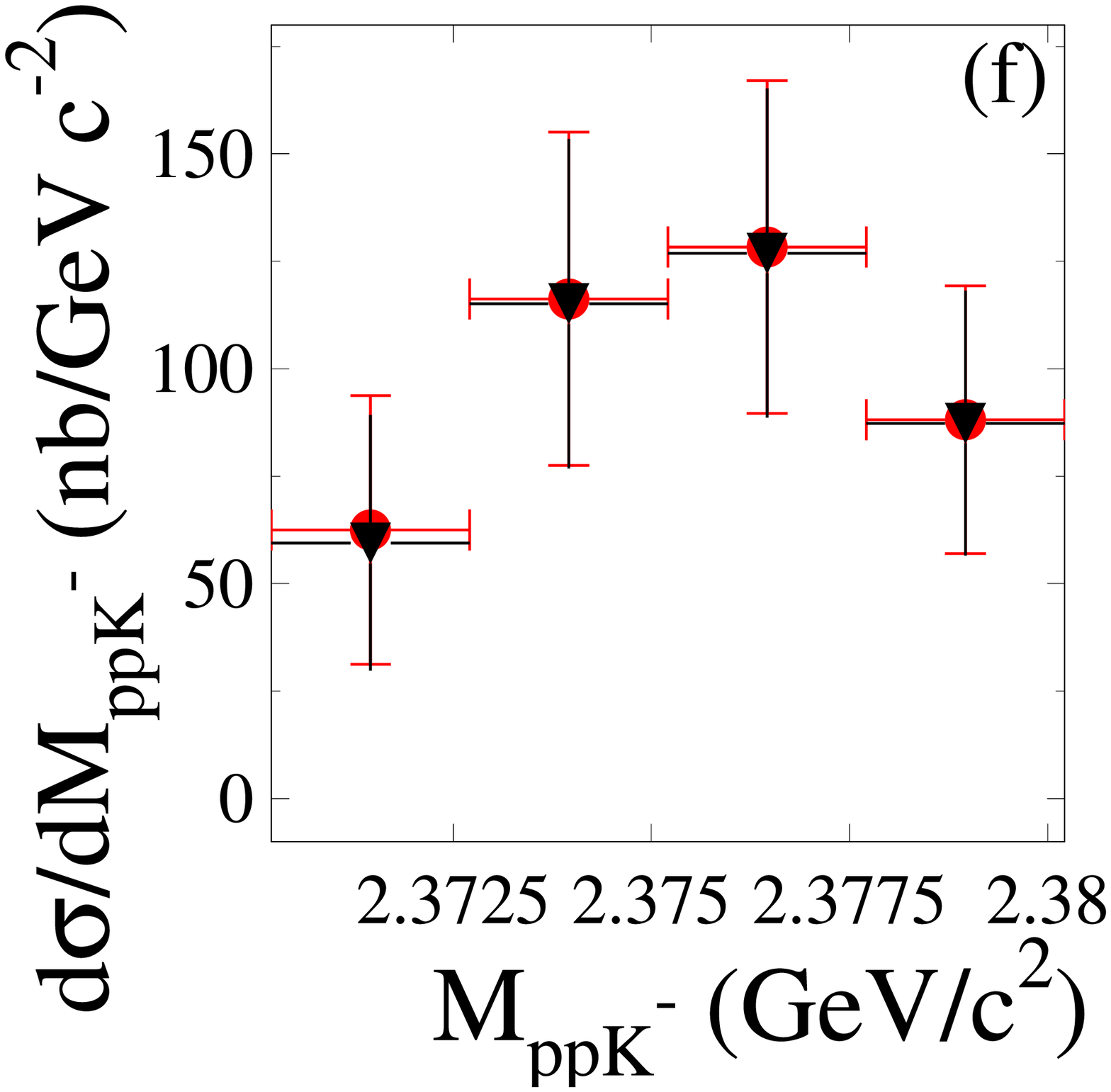}
\caption{(Color online) Differential cross sections for the $pp\rightarrow ppK^{+}K^{-}$ reaction
at Q~=~10~MeV. Circles and dashed bars (in black) denote spectra where the acceptance was determined taking
into account only the $pp$--FSI, and triangles  with solid bars (in red)
denote results where $pp$-- and $pK^{-}$--FSI was taken into account in the acceptance calculations.
They are hardly distinguishable. Vertical bars indicate statistical errors, whereas the horizontal
bars stand for the invariant mass intervals for which the cross section values were established.}
\label{diff10}
\end{figure}
The results are presented in Fig.~\ref{diff10} for data at Q~=~10~MeV and in
Fig.~\ref{diff28} for Q~=~28~MeV.
The distributions obtained under both assumptions are almost identical, which shows
that the acceptance of the COSY-11 detection setup is only very weakly sensitive to the
interaction between $K^{-}$ and protons. Thus, the observed enhancement in the excitation
function cannot be explained by approximations in the determination of the detection
efficiency as suspected by~\cite{anke}. This justifies the assumption made in the original
analysis, where the efficiency was calculated taking into account the $pp$--FSI only.\\
The derived values of differential cross sections are listed in Table~\ref{tab_all}.
This result constitutes an additional information to the total cross sections
published previously~\cite{winter}. The values of the cross sections
in the former analysis~\cite{winter} were determined using the total
number of events identified as a $pp\rightarrow ppK^{+}K^{-}$ reaction and
the total acceptance of the COSY-11.
Now after the determination of the absolute values for the differential distributions one can 
calculate the total cross sections in a less model dependent manner
regardless of the assumption of the $pp$--FSI.
\begin{figure}
\centering
\includegraphics[width=0.23\textwidth,angle=0]{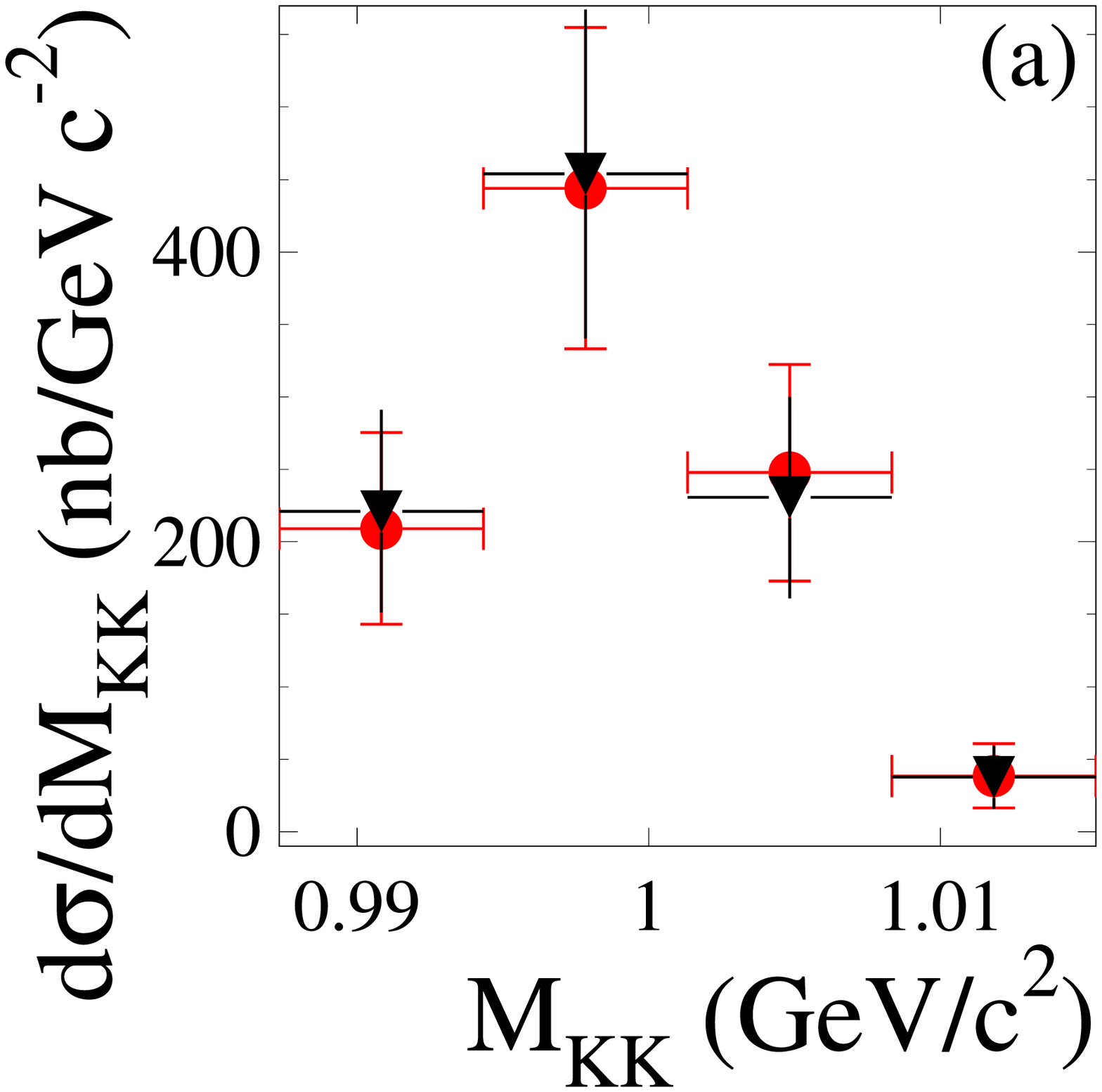}
\includegraphics[width=0.23\textwidth,angle=0]{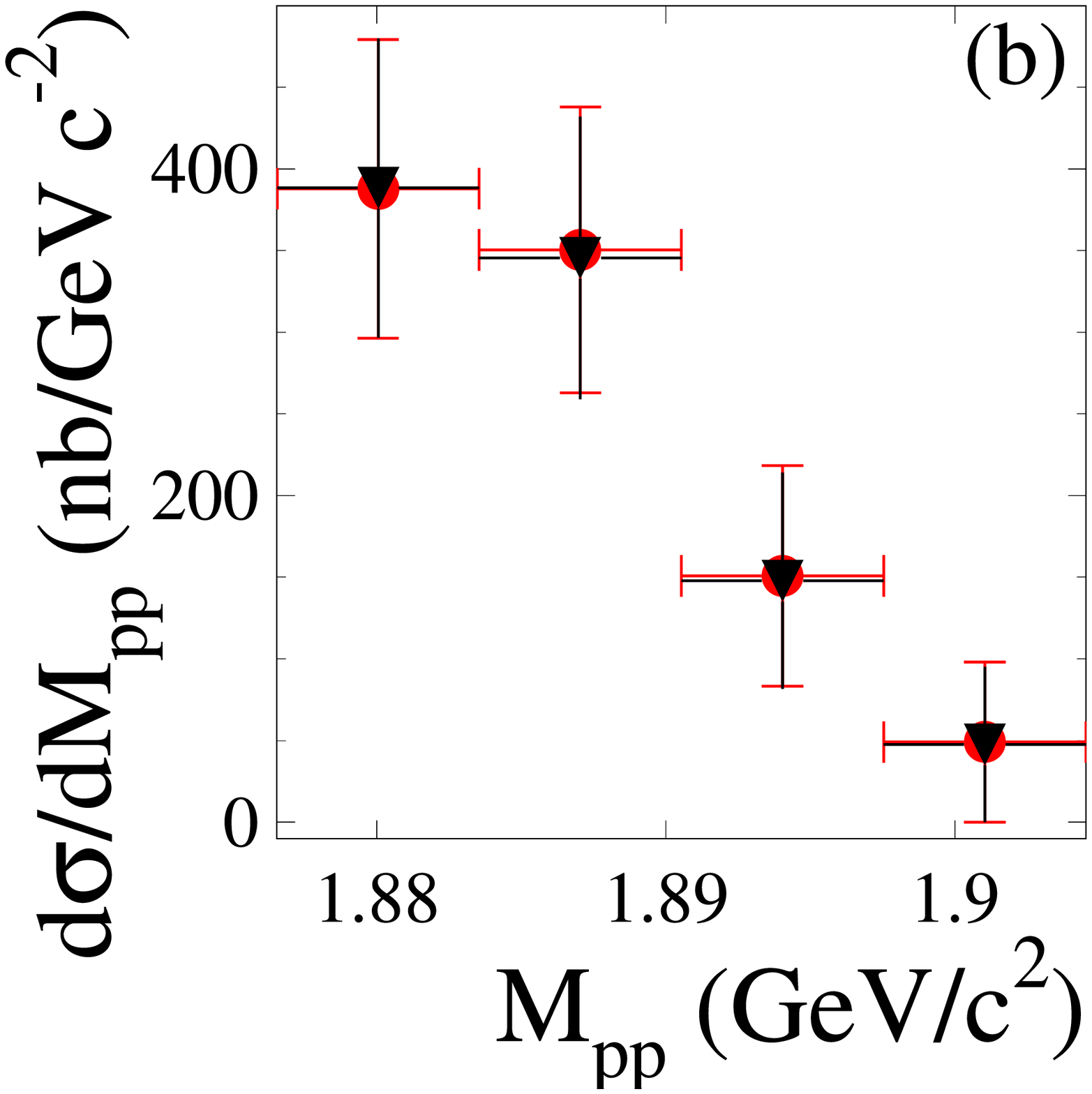}
\includegraphics[width=0.23\textwidth,angle=0]{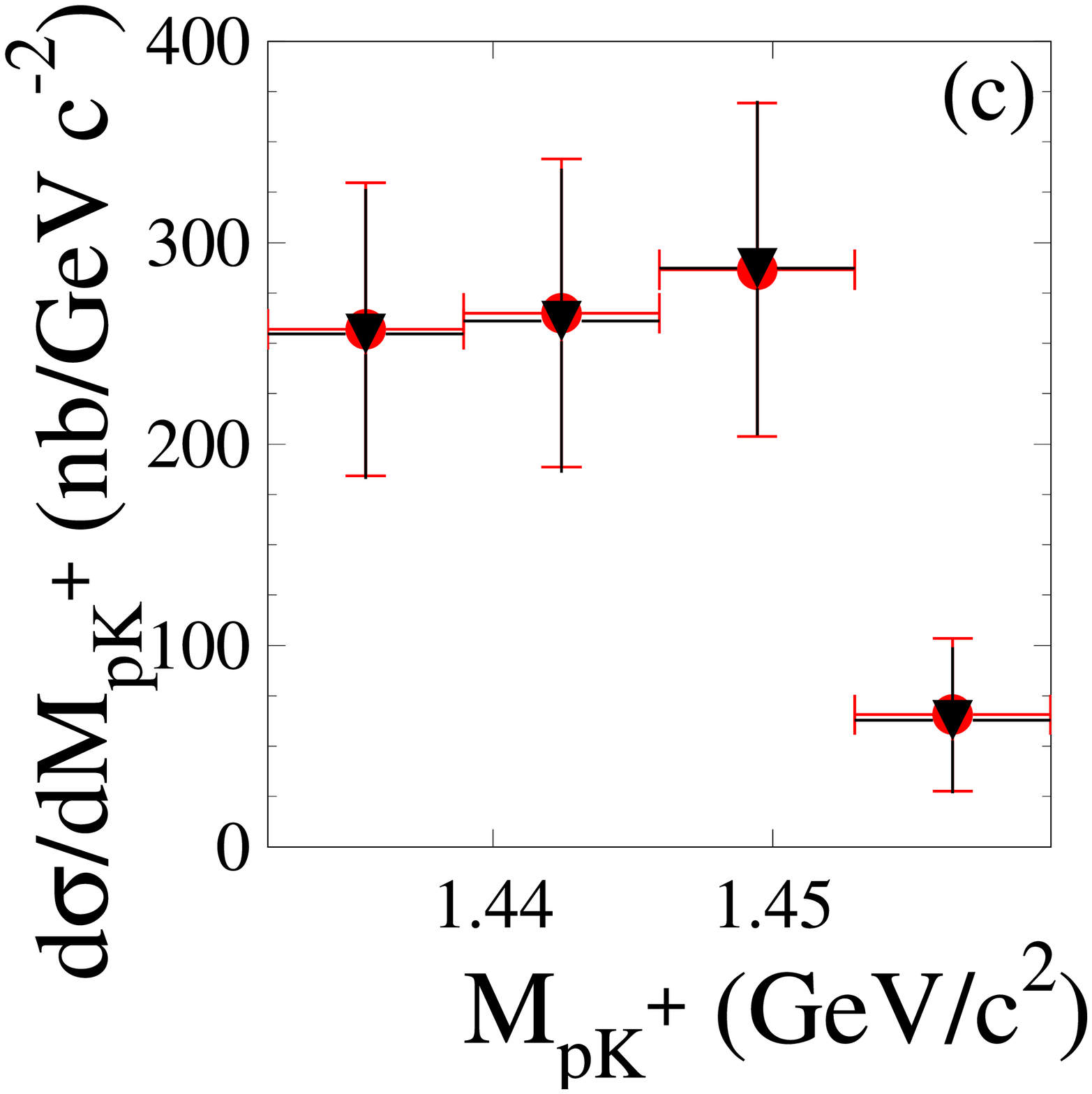}
\includegraphics[width=0.23\textwidth,angle=0]{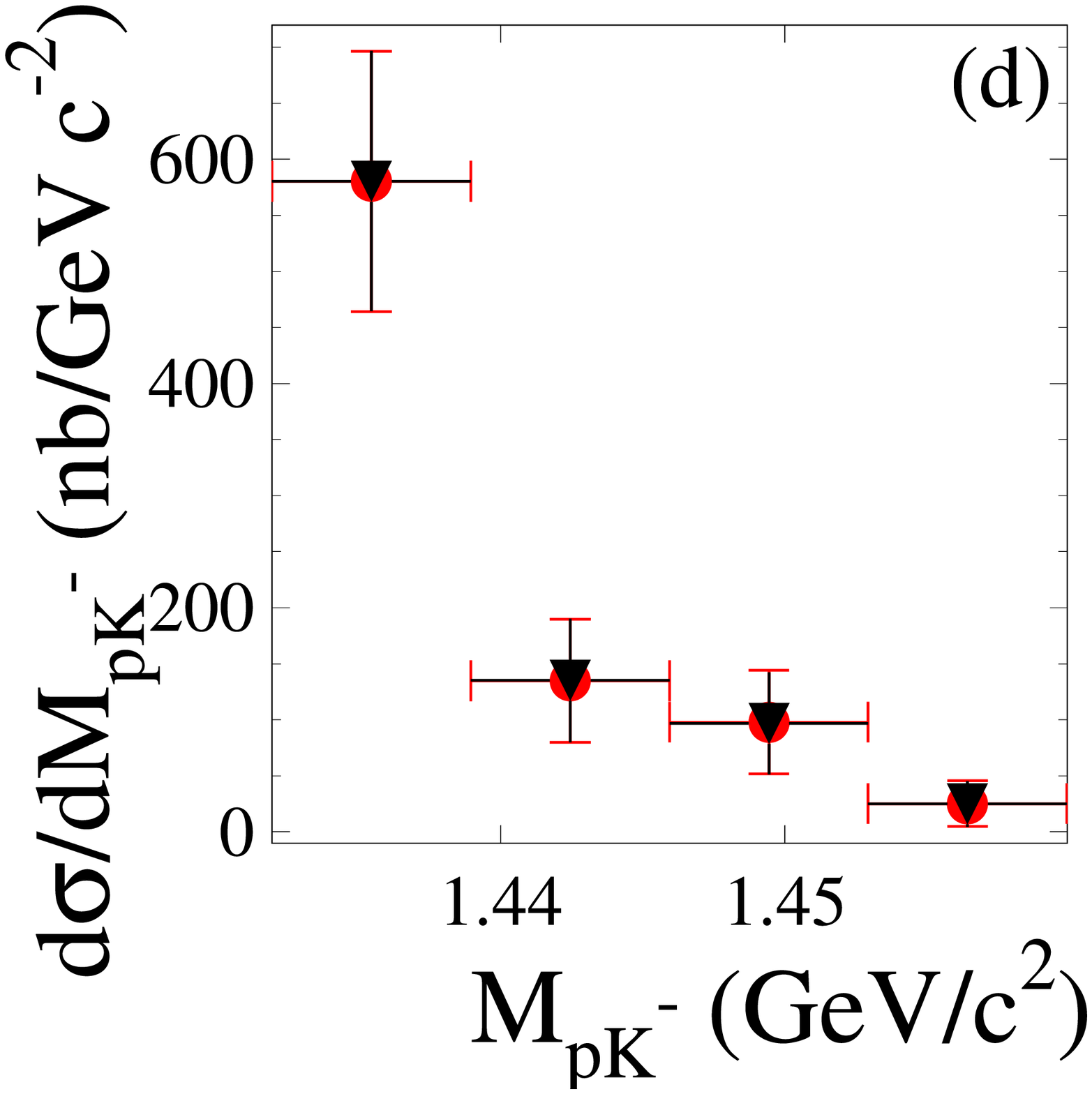}
\includegraphics[width=0.23\textwidth,angle=0]{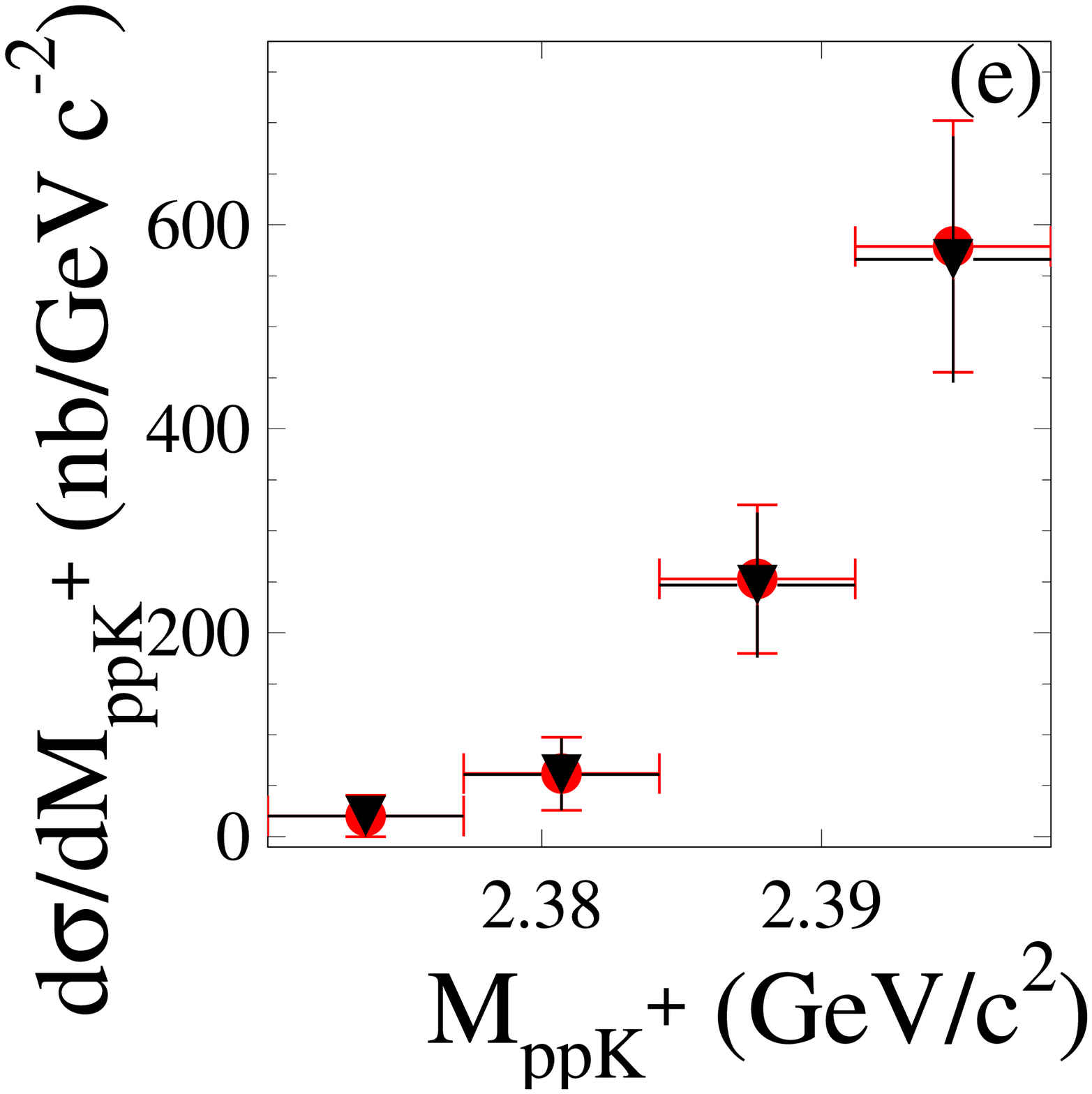}
\includegraphics[width=0.23\textwidth,angle=0]{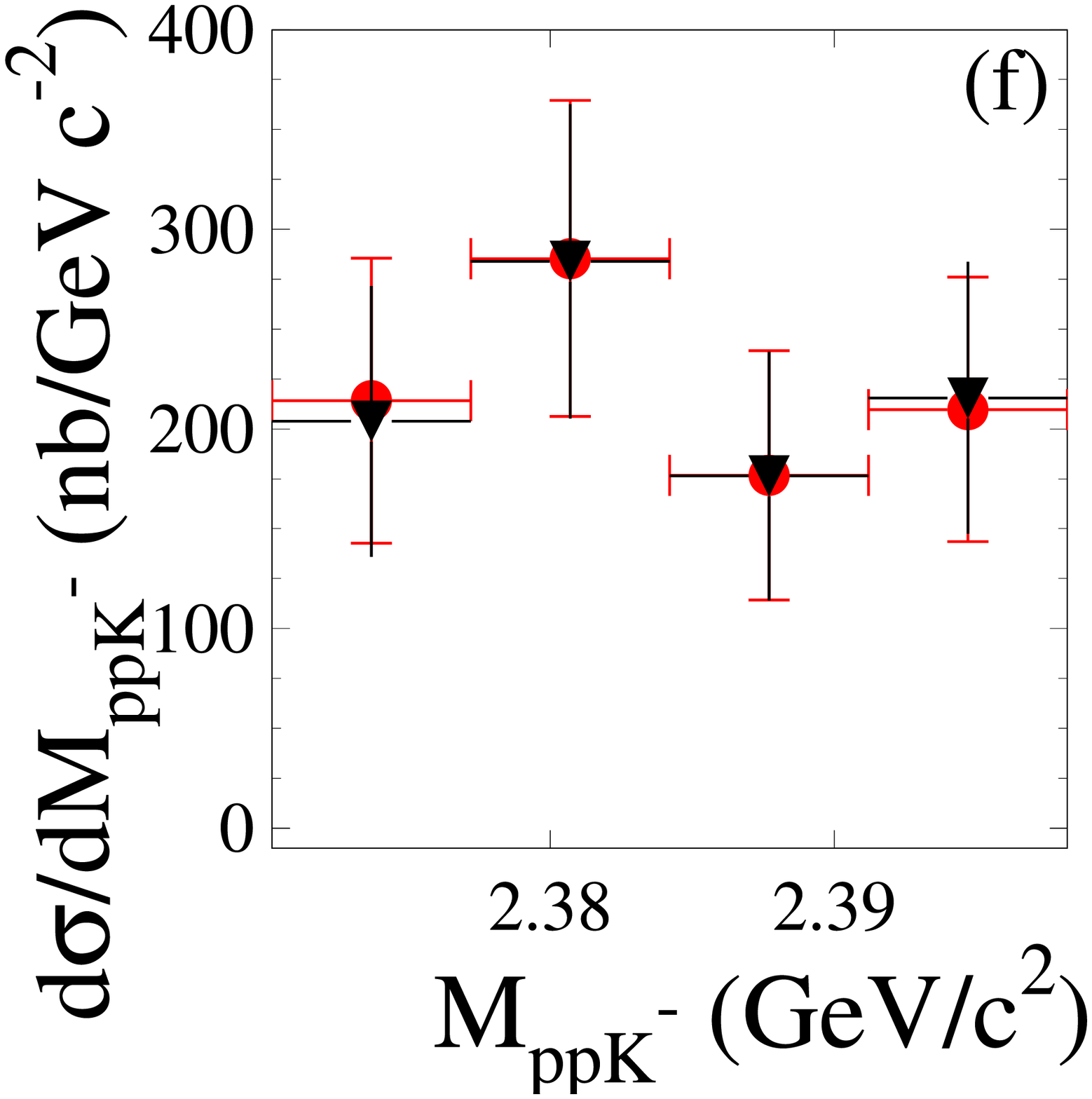}
\caption{(Color online) Differential cross sections for the $pp\rightarrow ppK^{+}K^{-}$ reaction
at Q~=~28~MeV. For the description see caption of Fig.~\ref{diff10}}
\label{diff28}
\end{figure}
%%%%%%%%
\begin{figure}
\centering
\includegraphics[width=0.23\textwidth,angle=0]{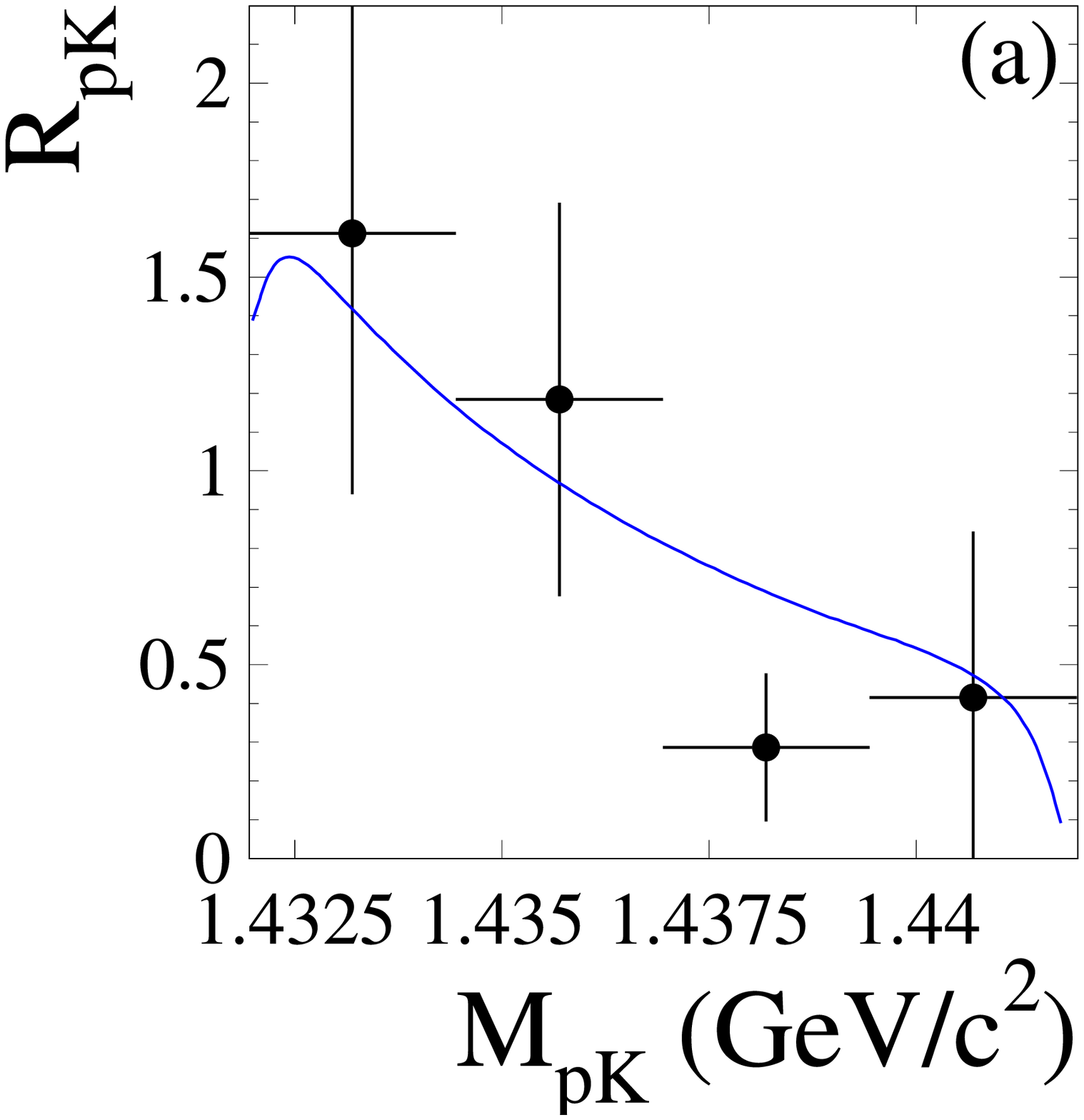}
\includegraphics[width=0.23\textwidth,angle=0]{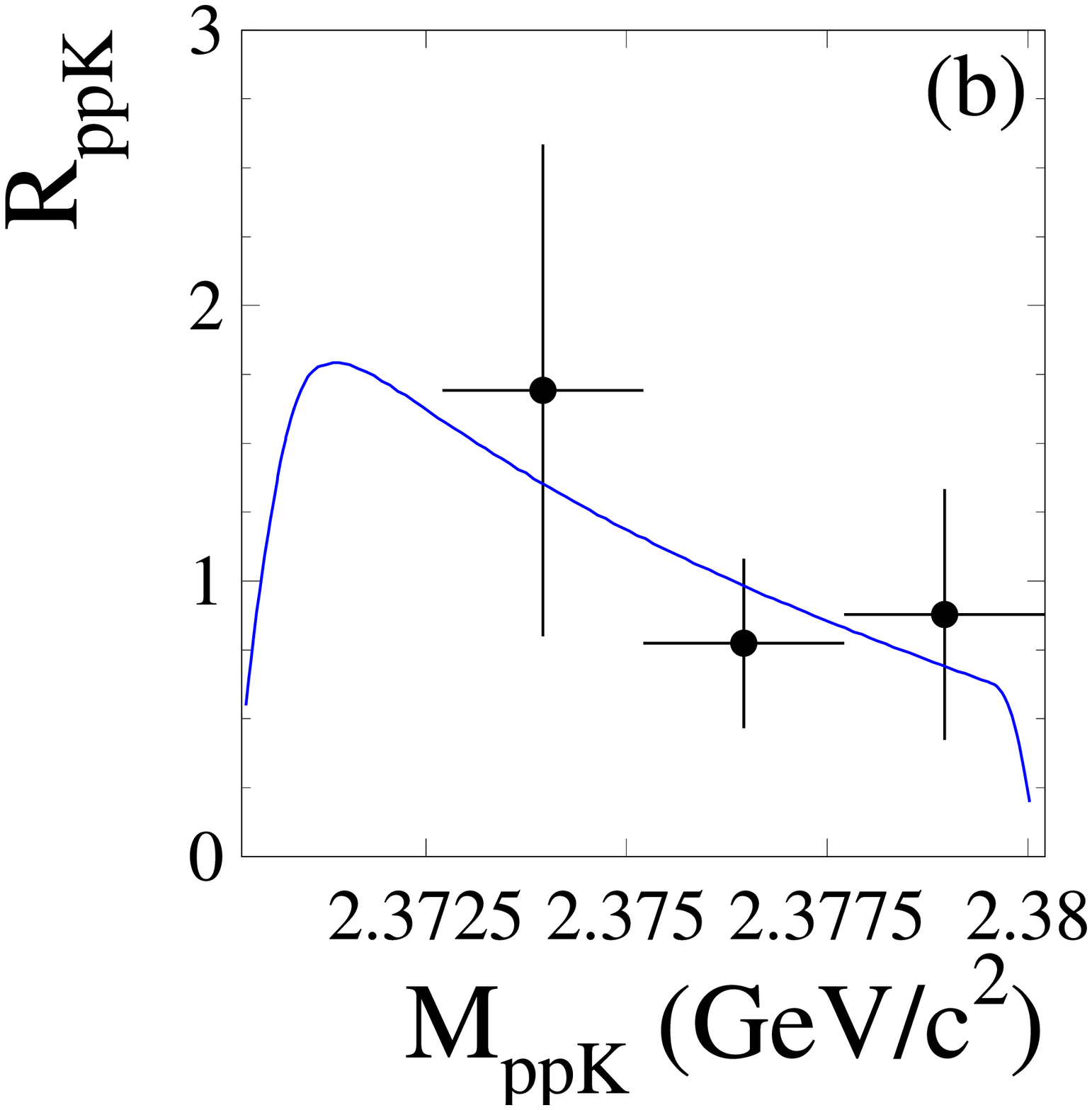}
\includegraphics[width=0.23\textwidth,angle=0]{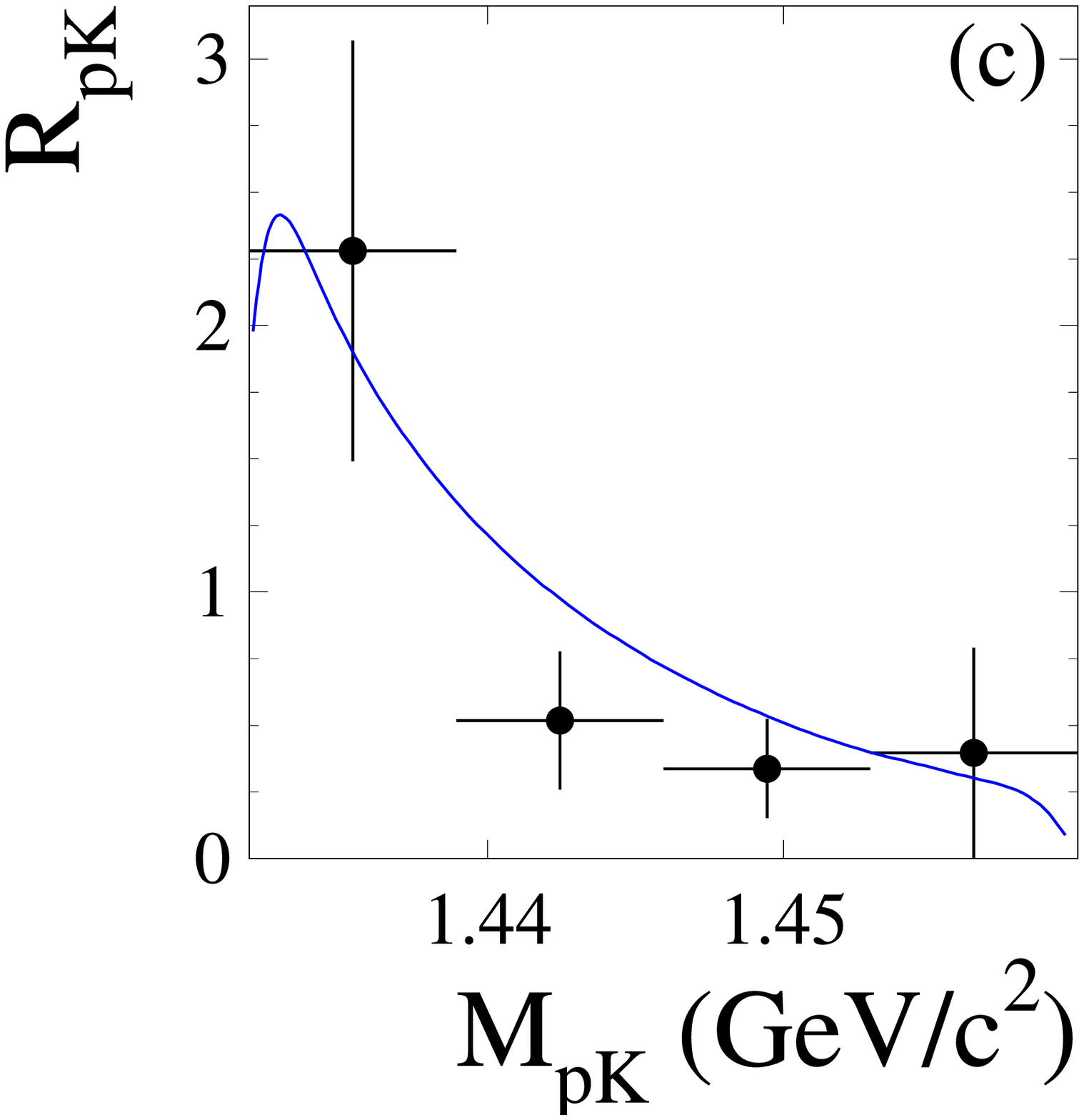}
\includegraphics[width=0.23\textwidth,angle=0]{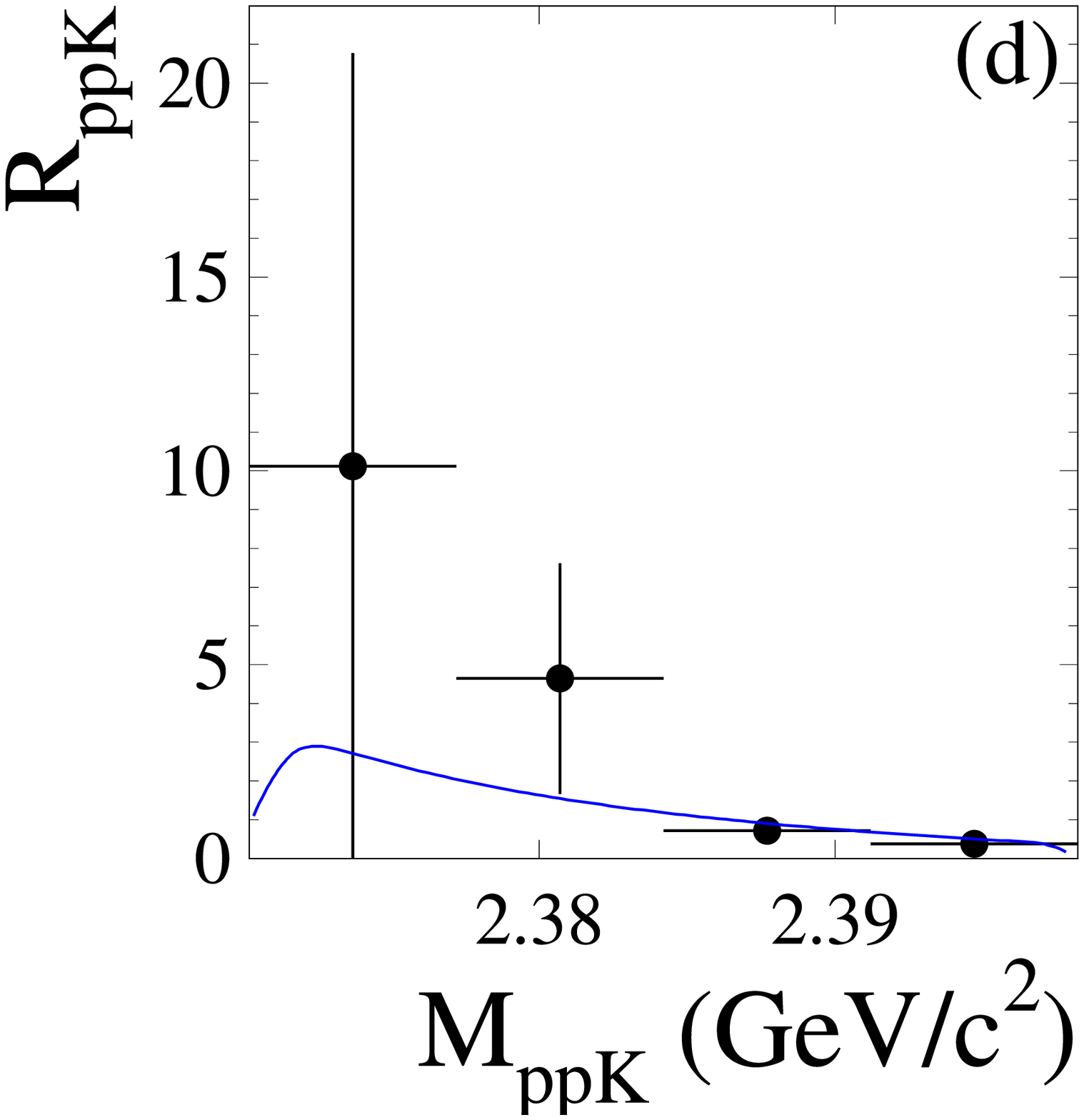}
\caption{(Color online) The distributions of the absolute 
values of ratios $R_{pK}$ and $R_{ppK}$
for data at Q~=~10~MeV ((a) and (b)) and Q~=~28~MeV ((c) and (d)).
Solid curves represent theoretical expectations calculated taking into account $pp$ and $pK^{-}$
final state interaction. It should be stressed that the absolute values are not fitted,
but they are result of the model and used parameters~\cite{anke}.}
\label{ratio}
\end{figure}
%%%%
The cross sections, calculated for both excess energies as an integral of the $M_{pp}$
distribution derived with the inclusion of the $pK^{-}$--FSI in the acceptance calculations:
\[\sigma_{\text{tot}}~=~\int\frac{d\sigma}{dM_{pp}}dM_{pp}~,\]
amount to $\sigma_{\text{tot}}~=~(0.95~\pm~0.17)~$nb for measurement at Q~=~10~MeV and
$\sigma_{\text{tot}}~=~(6.5~\pm~1.1)~$nb for Q~=~28~MeV.
These results are statistically consistent with the previously evaluated total cross sections.
However, the values extracted in the actual analysis with smaller error bars
are larger by about $20\%$ for Q~=~10~MeV and 50\% for Q~=~28~MeV,
which strengthens the confidence to the observed enhancement at threshold.\\
The determination of the absolute values for the differential cross sections
permitted us to establish the absolute values for the following ratios at the
close to threshold region~\footnote{In the former analysis only the shape of the ratios was established.}:
\begin{eqnarray*}
R_{pK}~=~\frac{d\sigma/dM_{pK^{-}}}{d\sigma/dM_{pK^{+}}}~,\\
R_{ppK}~=~\frac{d\sigma/dM_{ppK^{-}}}{d\sigma/dM_{ppK^{+}}}~.
\end{eqnarray*}
If $pK^+$ and $pK^-$ interactions were the same, the distribution of $R_{pK}$ as well as
$R_{ppK}$ should be flat and equal to unity. But as one can see in Fig.~\ref{ratio}
and as presented already in the previous publication by COSY-11~\cite{winter} and
ANKE~\cite{anke} for both excess energies $R_{pK}$ is far from being constant and
increases towards the lower $M_{pK}$ invariant masses. This effect might be due
to the influence of the $pK^{-}$ final state interaction. Similarly the distributions
of $R_{ppK}$ differs from expectations assuming only FSI in the $pp$ system.
This is a confirmation of effects observed also by the ANKE collaboration at
higher excess energies~\cite{anke}.
The determination of the absolute values of the ratios $R_{pK}$ and $R_{ppK}$
allows to compare the parameters of the scattering length $a_{pK^{-}}$
derived by the ANKE group (from data at excess energies above the $\phi$
meson production threshold), to the present data near the $K^+K^-$ threshold.
As demonstrated in Fig.~\ref{ratio}, simulations taking into account the
$pK^{-}$ final state interaction with the scattering length
$a_{pK^{-}}$~=~(0~+~1.5$i$)~fm determined
by the ANKE group from data at significantly higher excess energies reproduce
very well the distributions of $R_{pK}$ and $R_{ppK}$ near the threshold.
%############
\section{Analysis of the $K^{+}K^{-}$ final state interaction}
A factorization ansatz for the $pp$ and $pK^{-}$ interaction
underestimate the excitation function for the $pp\rightarrow ppK^{+}K^{-}$
reaction very close to threshold indicating that in this energy region
the influence of the $K^{+}K^{-}$ interaction is significant and cannot
be neglected. The interaction may manifest itself even stronger in the
distributions of the differential cross sections~\cite{review}.
This observation demands an analysis of the double differential cross sections
for the low energy data at Q~=~10~MeV (27~events) and Q~=~28~MeV (30~events),
in spite of the quite low statistics available.
\subsection{Generalization of the Dalitz plot: Goldhaber approach}
There are many different types of generalizations of the Dalitz plot
for four-body final states~\cite{nyborg,chodrow}.
Here a generalization proposed by Goldhaber is presented~\cite{goldhaber1,goldhaber2,wilkin2007}.

Consider a reaction like $a+b\longrightarrow 1+2+3+4$ yielding four particles with masses $m_{i}$
and total energy $\sqrt{s}$ in the centre-of-mass frame. The probability, that the momentum of
the $i^{th}$ particle is in a range $d^{3}p_{i}$ is given by:
\begin{eqnarray}
d^{12}P=d^{3}p_{1}d^{3}p_{2}d^{3}p_{3}d^{3}p_{4}\frac{1}{16E_{1}E_{2}E_{3}E_{4}}\nonumber\\
\times\delta^{3}\left(\sum_{j=1}^4 \mathbf{p_{j}}\right)\delta\left(\sum_{j=1}^4 E_{j}-\sqrt{s} \right)\left|M\right|^{2}~,
\label{prawdopodob}
\end{eqnarray}
where $E_{i}=\sqrt{\mathbf{p^2_i} + m^{2}_{i}}$ denotes the energy of the $i^{th}$
particle (c = 1) and $M$ denotes the invariant matrix element for the process.
Assuming that the matrix element $M$ depends only on the invariant masses
of the two- and three particle subsystems~\cite{nyborg}, the distribution given by Eq.~\ref{prawdopodob}
can be expressed in some choice of five independent invariant masses
\footnote
{Actually for four particles in the final state we have six two-particle and four three-particle invariant 
masses. But only five of them are independent due to following relations~\cite{nyborg}:
\[\sum_{i,j=1 (i>j)}^{4}\,M_{ij}^{2} = s + 2 \sum_{i=1}^{4}\,m_{i}^{2}~,\]
\[M_{123}^{2} = M_{12}^{2} + M_{23}^{2} + M_{13}^{2} - m_{1}^{2} - m_{2}^{2} - m_{3}^{2}~,\]
\[M_{134}^{2} = M_{13}^{2} + M_{34}^{2} + M_{14}^{2} - m_{1}^{2} - m_{3}^{2} - m_{4}^{2}~,\]
\[M_{124}^{2} = M_{12}^{2} + M_{24}^{2} + M_{14}^{2} - m_{1}^{2} - m_{2}^{2} - m_{4}^{2}~,\]
\[M_{234}^{2} = M_{22}^{2} + M_{34}^{2} + M_{24}^{2} - m_{2}^{2} - m_{3}^{2} - m_{4}^{2}~.\] 
}. An especially convenient choice is
$M_{12}^{2},M_{34}^{2},M_{14}^{2},M_{124}^{2},$ and $M_{134}^{2}$~\cite{nyborg}.
Using such variables, after integrations, one obtains an event distribution in the following form:
\begin{equation}
d^{5}P=\frac{\pi^{2}}{8s}\left|M\right|^{2}\frac{1}{\sqrt{-B}}~dM^{2}_{12}dM^{2}_{34}dM^{2}_{14}dM^{2}_{124}dM^{2}_{134}~,
\label{goldhaber3}
\end{equation}
where $B$ is a function of the invariant masses with the exact form to be found in Nyborg's work~\cite{nyborg}.
An advantage of the choice of mass variables used here is the high symmetry of the function $B$~\cite{nyborg}, which is very
usefull in the consideration of the boundary of the kinematically allowed region in the $\left(M_{12}^{2},M_{34}^{2},M_{14}^{2},
M_{124}^{2},M_{134}^{2}\right)$ space defined by B~=~0.
If we additionally change the integration variables to invariant masses the projection of
the physical region on the $\left(M_{12},M_{34}\right)$-plane gives a convenient and simple
distribution, which can be used to analyse the final state interaction in the same way as
in case of three particles. Such an analysis was originally made by Goldhaber \textit{et al.}
in 1963~\cite{goldhaber1,goldhaber2} and is called Goldhaber plot.
As it is shown in Fig.~\ref{goldhabery} the kinematically allowed region in
the Goldhaber plot is a right isosceles triangle within which the area (contrary to the Dalitz plot) is not proportional
to the phase space volume~\cite{nyborg}. Moreover, even
in the case of the absence of any final state interaction
the density of events on the Goldhaber plot is not homogeneous. However, it is still a very
convenient tool due to its lorentz invariance and simple boundary equations~\cite{nyborg}:
\[M_{12}+M_{34} = \sqrt{s}~,~M_{12} = m_{1}+m_{2}~, \text{and}~M_{34} = m_{3}+m_{4}~.\]
%####
\subsection{Determination of the $K^{+}K^{-}$ scattering length}
\begin{figure}
\centering
\includegraphics[width=0.23\textwidth,angle=0]{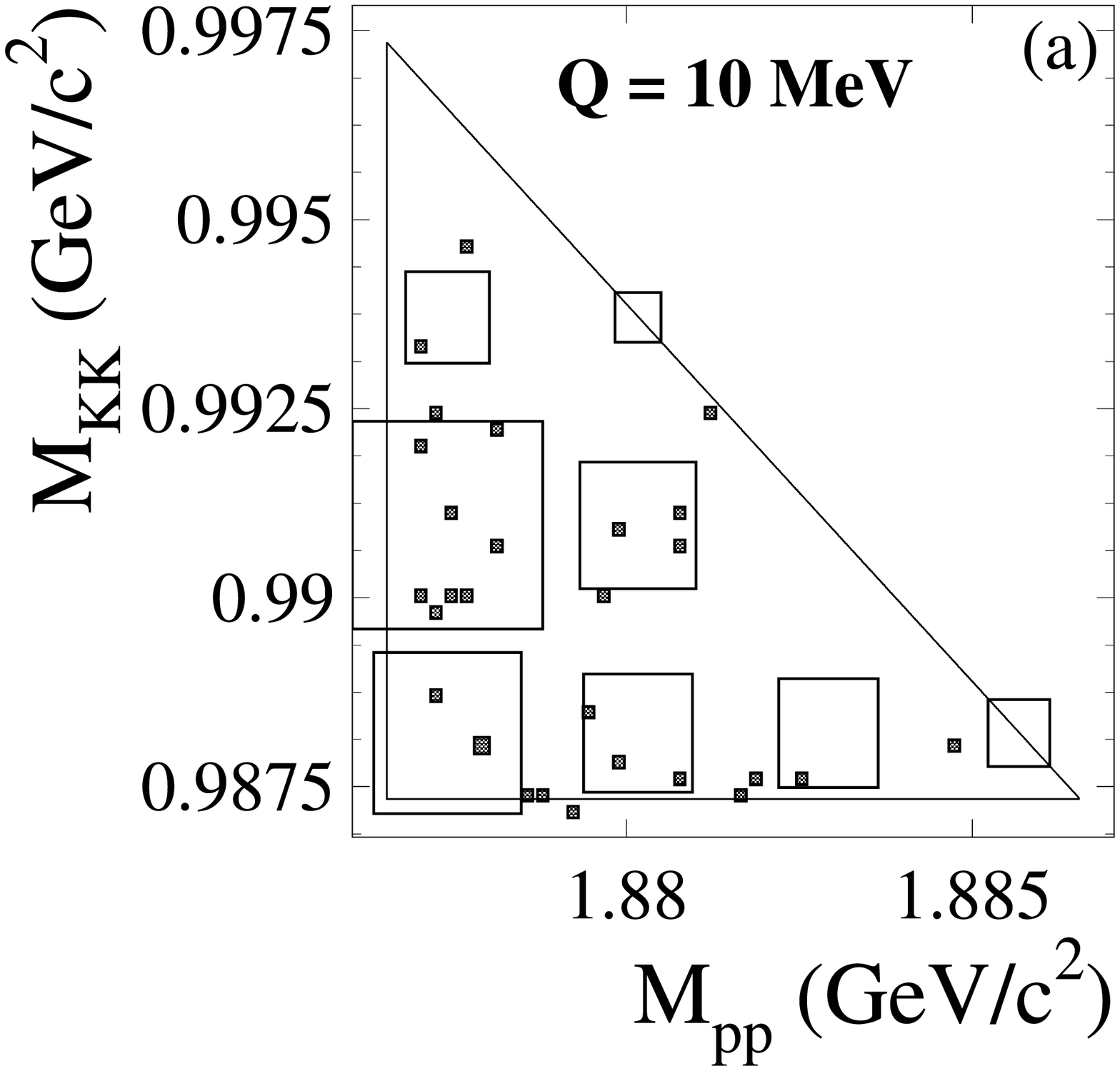}
\includegraphics[width=0.23\textwidth,angle=0]{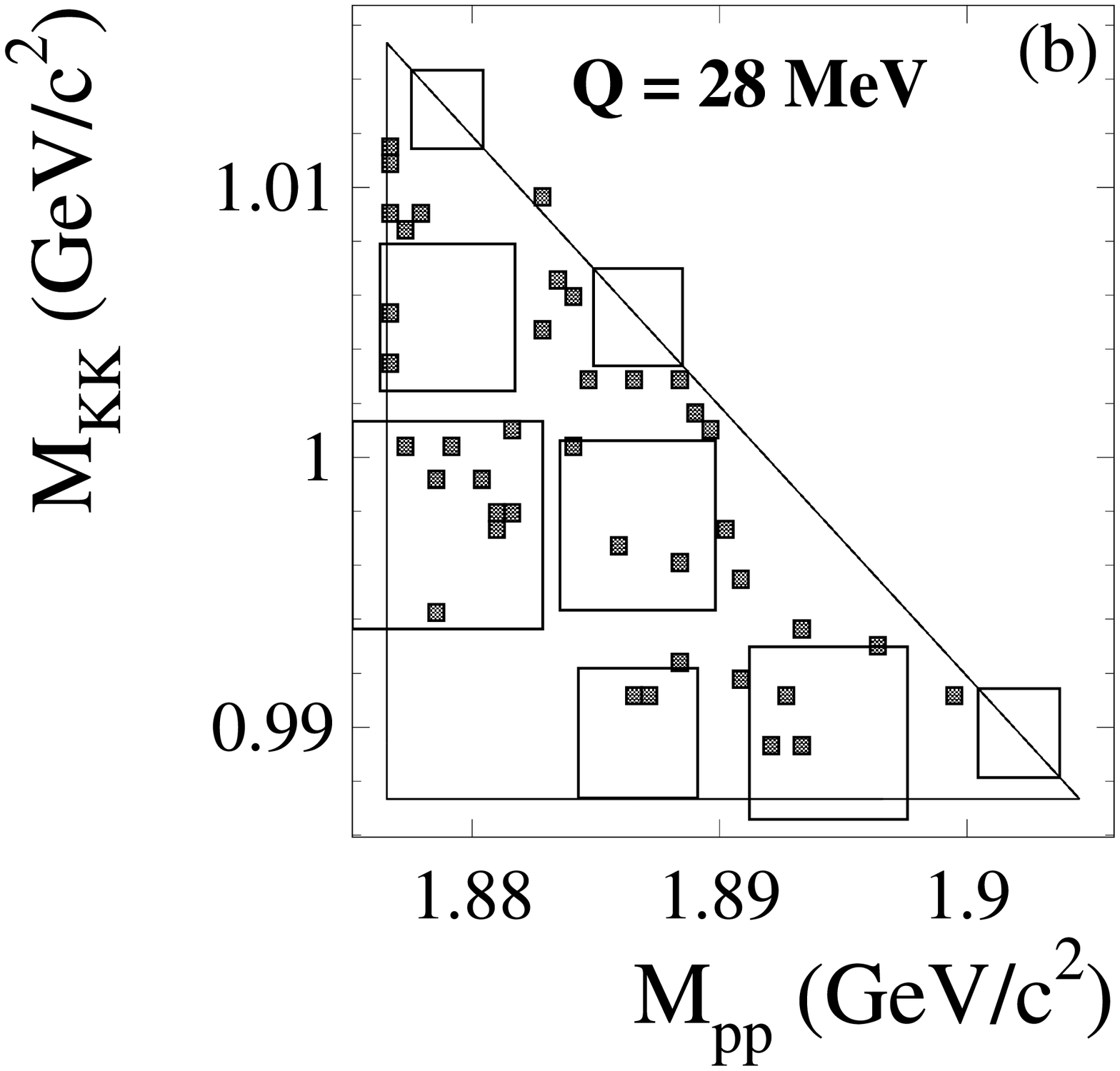}
\includegraphics[width=0.23\textwidth,angle=0]{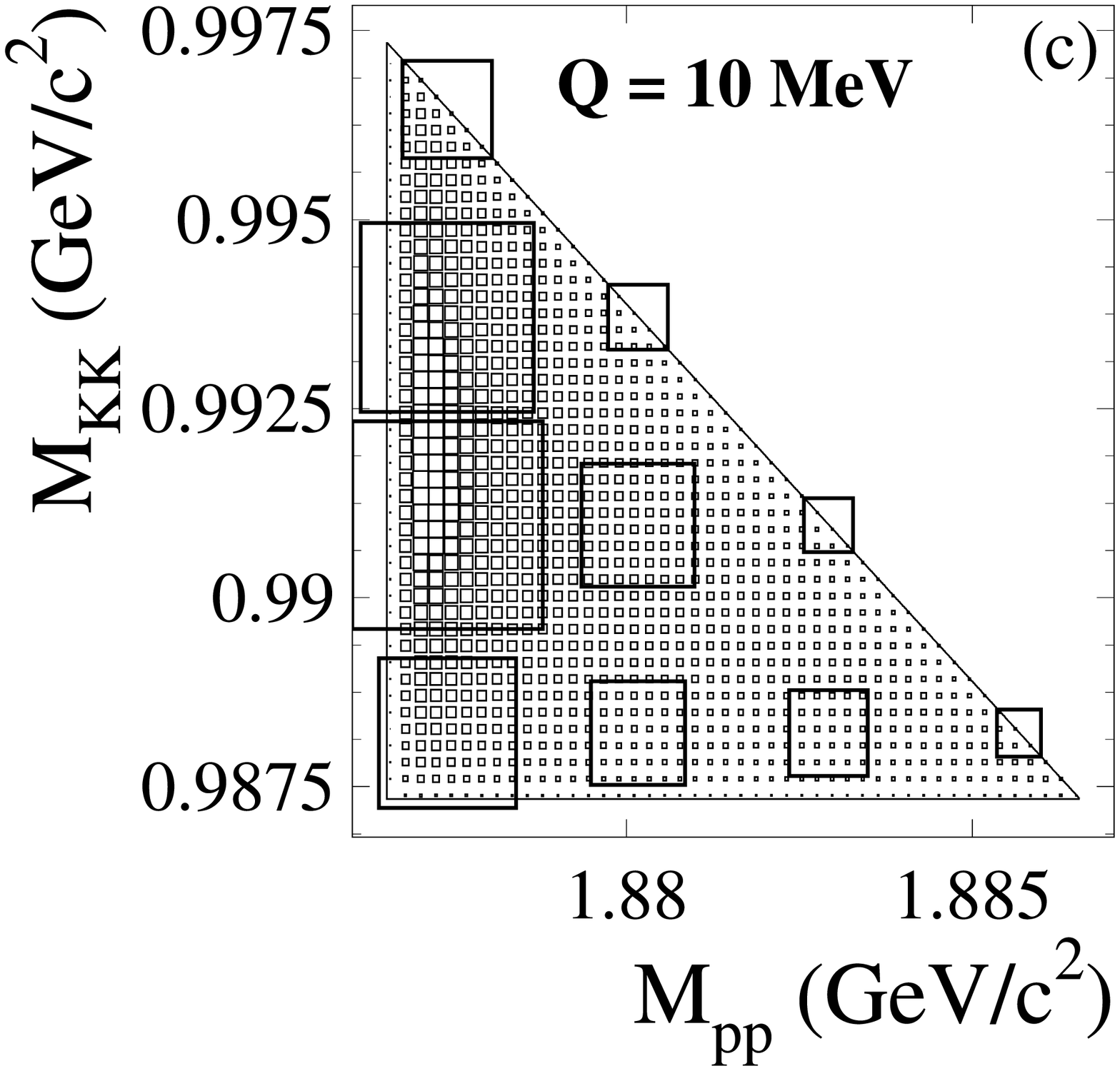}
\includegraphics[width=0.23\textwidth,angle=0]{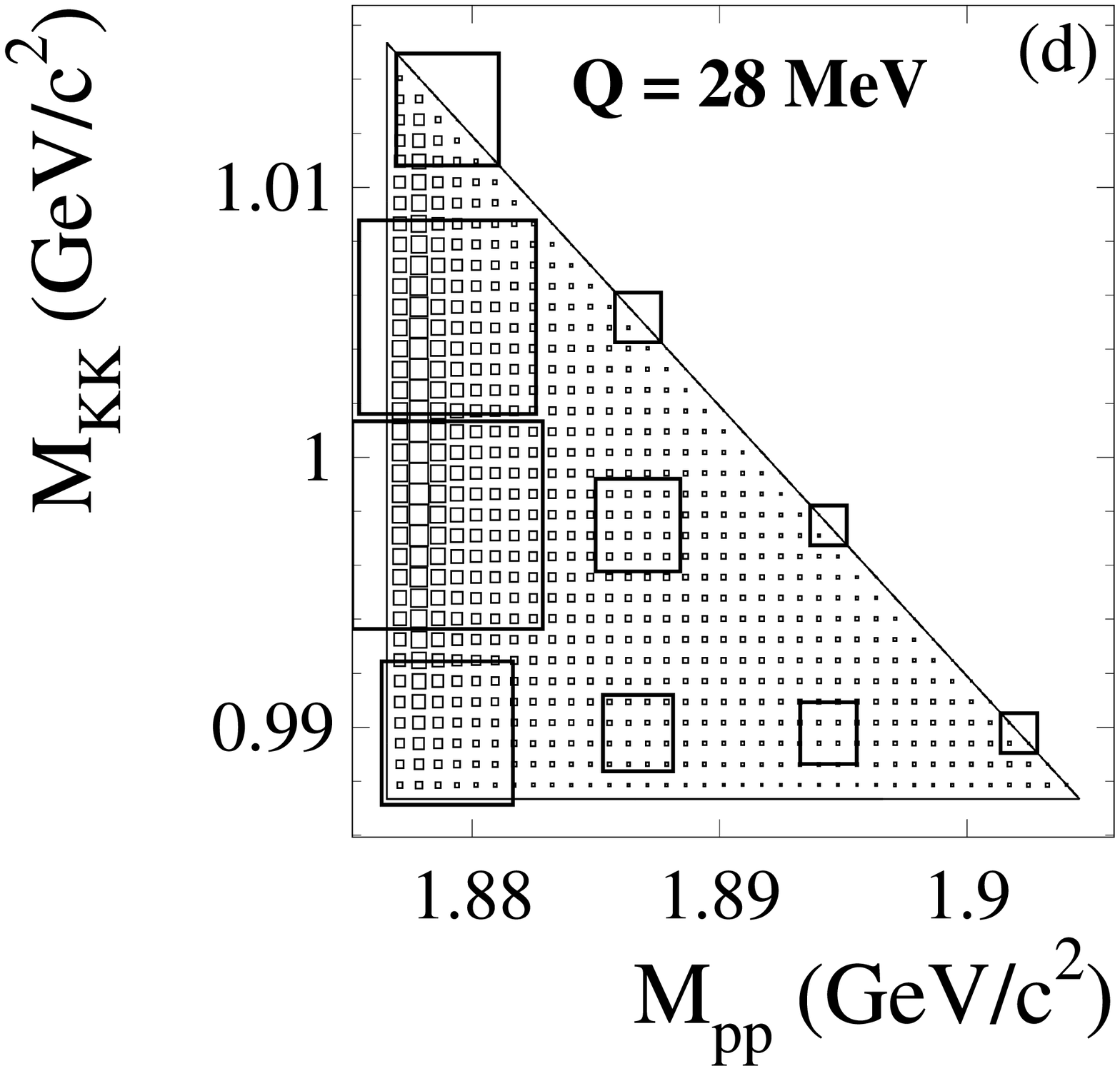}
\caption{
Goldhaber plots for the $pp\rightarrow ppK^{+}K^{-}$ reaction. 
The solid lines of the triangles show the 
kinematically allowed boundaries. 
Raw data are shown in Figs. (a) and (b)  as black points. 
The superimposed squares represent the same distributions but binned into intervals of
$\Delta$M~=~2.5~MeV/$\text{c}^{2}$ ($\Delta$M~=~7~MeV/$\text{c}^{2}$) widths for an excess energy 
of Q~=~10~MeV (28~MeV), respectively.
The size of the square is proportional to the number of entries in a given interval.
In Figs. (c) and (d)  Monte Carlo
results are presented. 
In the simulated distributions both the $pp$ and the $pK^-$--FSI are taken into account.
 }
\label{goldhabery}
\end{figure}
Complementary to previous derivations~\cite{kaminski,Baru,Teige,Bugg} here we estimate the $K^{+}K^{-}$
scattering length directly from the low energy differential mass distributions of $K^{+}K^{-}$ and pp pairs
from the $ppK^{+}K^{-}$ system produced at threshold.
The raw data (represented by black points in Figs.~\ref{goldhabery}(a) and ~\ref{goldhabery}(b)) were first binned into intervals of $\Delta$M~=~2.5~MeV/$\text{c}^{2}$ width for
the measurement at Q~=~10~MeV and intervals of 
$\Delta$M~=~7~MeV/$\text{c}^{2}$ for the data at Q~=~28~MeV,
and then for each bin corrected for the acceptance 
and detection efficiency of the COSY-11 facility~\cite{mich_mgr}.
The resulting Goldhaber plots are presented 
together with the raw distributions 
in Figs.~\ref{goldhabery}(a) and \ref{goldhabery}(b). 
Figures~\ref{goldhabery}(c) and \ref{goldhabery}(d) show corresponding distributions simulated
with Monte Carlo method taking into account the $pp$ and $pK^-$
interaction according to the factorization ansatz~\cite{anke}.

In order to estimate the strength of the $K^+K^-$ interaction,
the derived cross sections were compared to results of simulations
generated with various parameters of the $K^{+}K^{-}$ 
interaction taking into account strong final state interaction
in the $pp$ and $pK^{-}$ subsystems.
To describe the experimental data in terms of final state interactions between
i) the two protons, ii) the $K^-$ and  protons and iii) the $K^{+}$ and $K^{-}$, the $K^{+}K^{-}$ enhancement
factor was introduced such that Eq.~\ref{pp-pkfsi} changes to:
\begin{equation}
F_{FSI}~=~F_{pp}(q)\cdot F_{p_{1}K^{-}}(k_{1})\cdot F_{p_{2}K^{-}}(k_{2})\cdot F_{K^{+}K^{-}}(k_{3})~.
\label{pp-pk-kk_fsi}
\end{equation}
As for the case of the $pK^{-}$--FSI, the $F_{K^+K^-}$ was calculated in the scattering length approximation:
\begin{equation}
F_{K^{+}K^{-}}~=~\frac{1}{1~-~i~k_{3}~a_{K^+K^-}}~,
\label{F_KK}
\end{equation}
where $a_{K^{+}K^{-}}$ is the effective $K^{+}K^{-}$ scattering length and $k_{3}$ stands for the relative momentum of
 the kaons in their rest frame.
\begin{figure}
\includegraphics[width=0.2386\textwidth,angle=0]{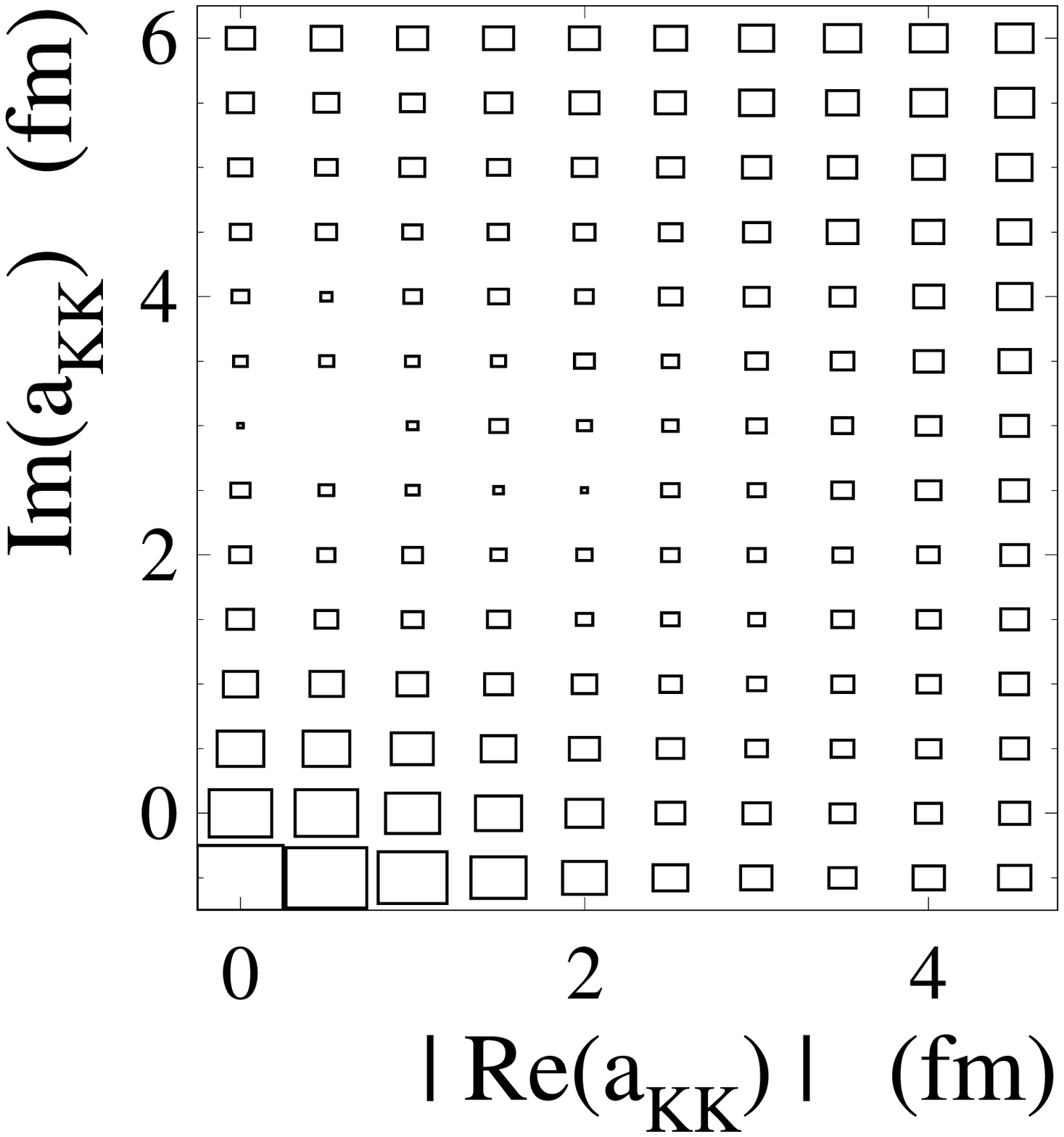}
\includegraphics[width=0.2386\textwidth,angle=0]{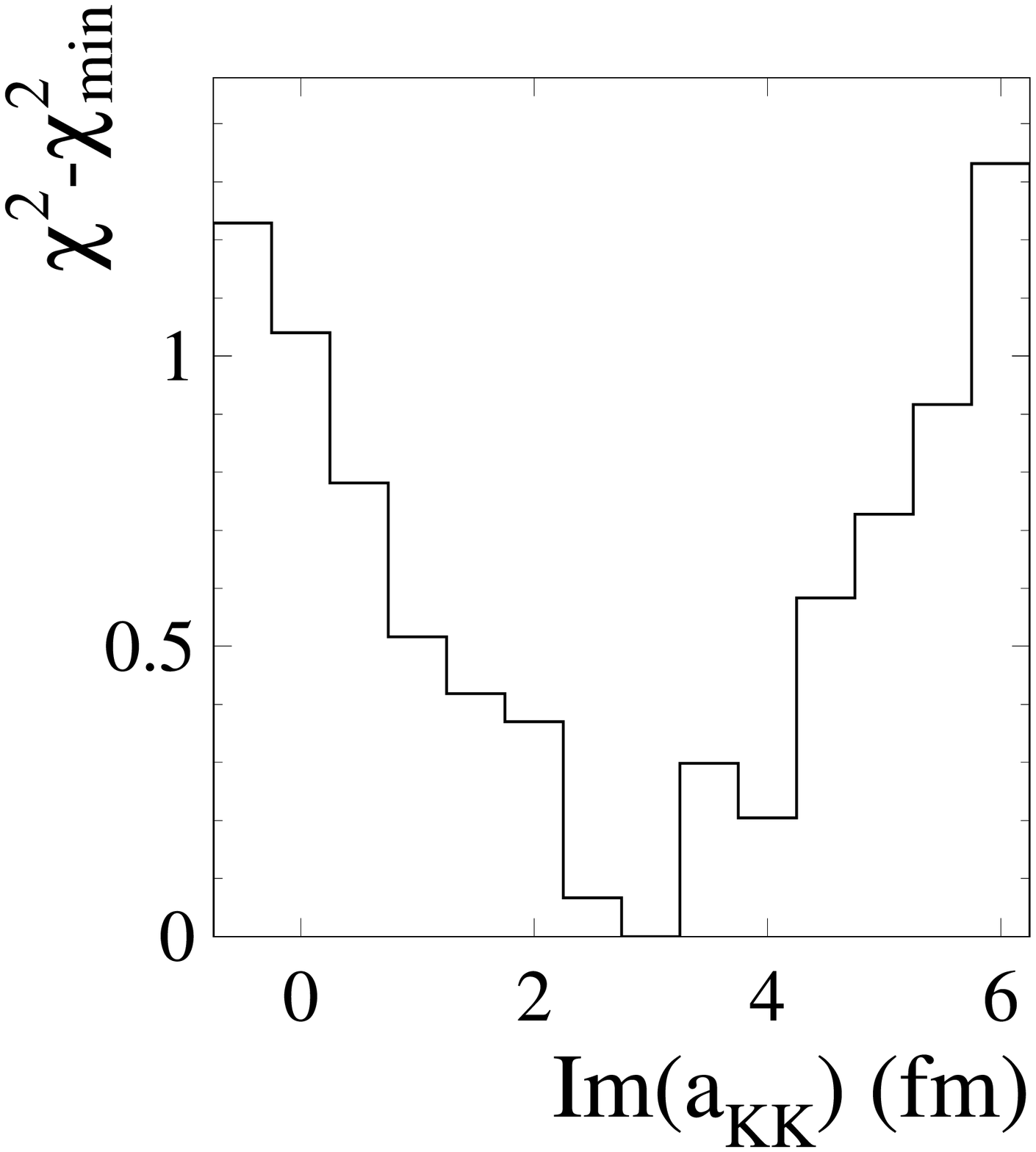}
\includegraphics[width=0.2386\textwidth,angle=0]{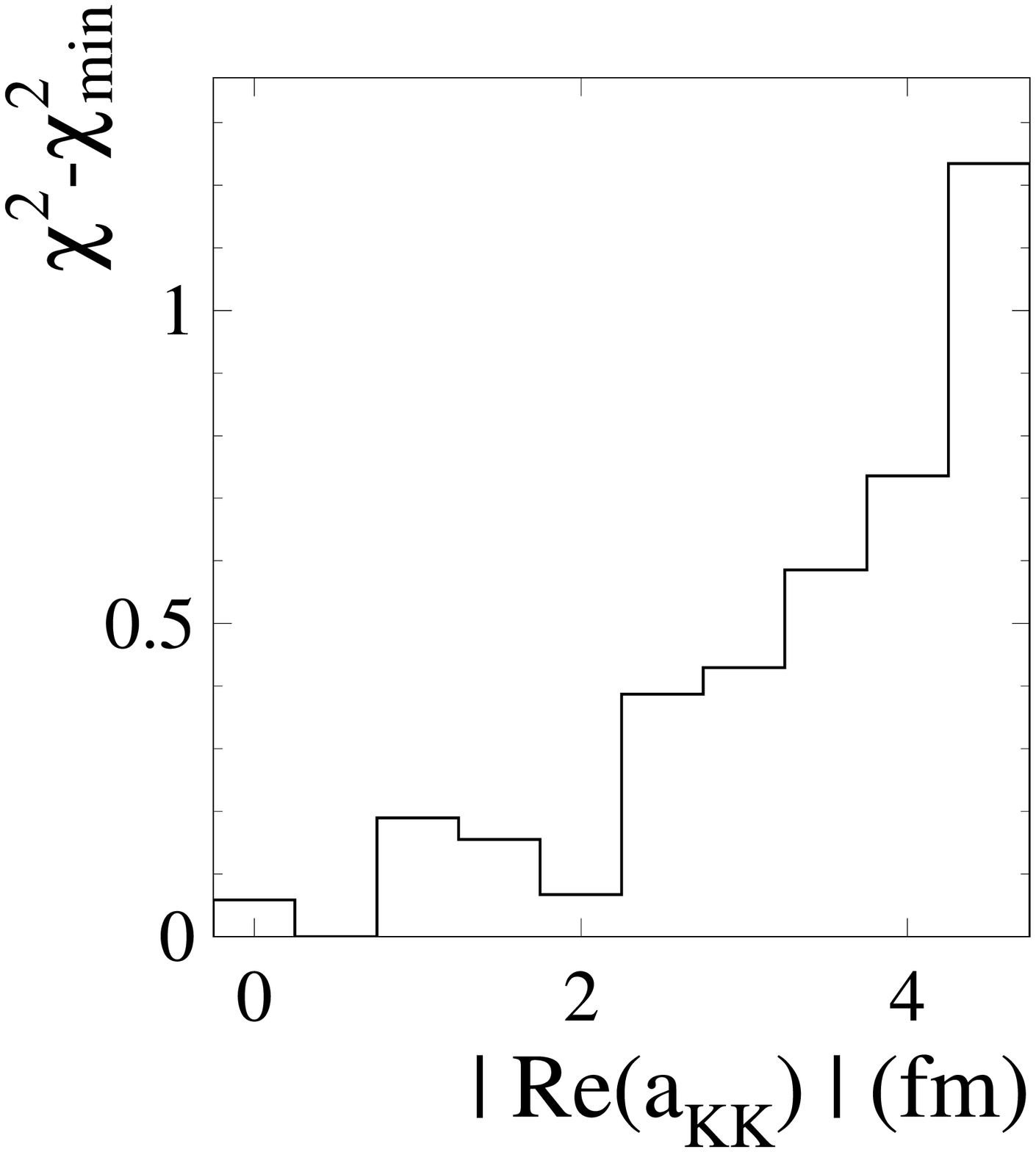}
\caption{
$\chi^{2}$~-~$\chi^{2}_{min}$ distribution as a function of $|Re(a_{K^{+}K^{-}})|$ and $Im(a_{K^{+}K^{-}})$.
$\chi^{2}_{min}$ denotes the absolute minimum with respect to parameters $\alpha$, $|Re(a_{K^{+}K^{-}})|$,
and $Im(a_{K^+K^-})$.}
 \label{chi2_1}
\end{figure}
Using this parametrization we compared the experimental event distributions
to the results of Monte Carlo simulations treating the $K^+K^-$ scattering length as an unknown parameter, which
has to be determined. In order to estimate the real and imaginary part of
$a_{K^{+}K^{-}}$ we constructed the Poisson likelihood $\chi^{2}$ statistic derived from the
maximum likelihood method~\cite{bakernim,feldmanpr}:
\begin{equation}
\chi^2\left(a_{K^+K^-},\alpha\right) = 2 \cdot \sum_i \, [\alpha N_i^s - N_i^e +  N_i^e \,
\text{ln}(\frac{N_i^e}{\alpha N_i^s})]~,
\label{eqchi2_mh}
\end{equation}
where $N_i^e$ denotes the number of events in the $i^{th}$ bin of the experimental Goldhaber plot,
$N_i^s$ stands for the content of the same bin in the simulated distributions, and $\alpha$
is the normalization factor.
The data collected at both excess energies have been analysed simultaneously.
The obtained $\chi^2$ distributions (suppressed by its minimum value) as a function of the real
and imaginary part of the $K^+K^-$ scattering length are presented in Fig.~\ref{chi2_1}.
The best fit to the experimental data corresponds
to $\left|Re(a_{K^{+}K^{-}})\right| = 0.5^{~+4}_{~-0.5}$~fm and $Im(a_{K^{+}K^{-}}) = 3~\pm~3$~fm.
The final state interaction enhancement factor $F_{K^{+}K^{-}}$ in the scattering length approximation is symmetrical
with respect to the sign of $Re(a_{K^{+}K^{-}})$, therefore only its absolute value can be determined.
%###########
\section{Conclusions}
In conclusion, the more detailed analysis of the COSY-11 data with inclusion of the $pK^-$ final state
interaction did not change significantly the result of the previous analysis~\cite{winter}.
Moreover, the new more precise determination of the total cross sections from the differential
$M_{pp}$ distributions even increased the enhancement at threshold.\\
In addition the analysis of the $pp\rightarrow ppK^{+}K^{-}$ reaction has been extended to the
determination of differential cross sections in view of the $K^{+}K^{-}$ final state interaction.
The extracted $K^+K^-$ scattering length amounts to:
\begin{center}
$a_{K^+K^-}$~=~[(0.5$^{~+4}_{~-0.5}$)~+~$i$(3~$\pm$~3)]~fm.
\end{center}
Due to the low statistics the uncertainties are rather large.
In this analysis we cannot distinguish between the isospin I~=~0 and I~=~1 states of the $K^+K^-$ system.
However, as pointed out in~\cite{dzyuba}, the production with I~=~0 is dominant in the $pp\to ppK^+K^-$ reaction
independent of the exact values of the scattering lengths.

Regarding  the comparison of the interactions in the $pK^{-}$, $pK^{+}$, $ppK^{-}$ and
$ppK^{+}$ subsystems, the absolute ratios determined from the COSY-11 data measured at
Q~=~10~MeV and Q~=~28~MeV are consistent with the predictions based on the parametrization
introduced in reference~\cite{anke} and on the values of the scattering length $a_{pK^-}$
extracted from the ANKE data at higher excess energies~\cite{anke}.
%#############
\begin{acknowledgments}
The work was
supported by the
European Community-Research Infrastructure Activity
under the FP6 program (Hadron Physics,RII3-CT-2004-506078), by
the German Research Foundation (DFG), by
the Polish Ministry of Science and Higher Education through grants
No. 3240/H03/2006/31  and 1202/DFG/2007/03,
and by the FFE grants from the Research Center J{\"u}lich.
\end{acknowledgments}


\begin{thebibliography}{9}
\bibitem{oelert}
W.~Oelert, \textit{in Proceedings of the Workshop on Meson Production, Interaction and Decay,
  Cracow, 1991,} World Scientific, Singapore (1991) p. 199.
%\href{http://www.slac.stanford.edu/spires/find/hep/www?irn=4799836}{SPIRES entry}
\bibitem{Morgan}
D.~Morgan,~M.~R.~Pennington,~Phys. Rev.~D~\textbf{48}, 1185 (1993).
%%CITATION = PHRVA,D48,1185;%%
\bibitem{Jaffe}
R.~L.~Jaffe,~Phys. Rev.~D \textbf{15}, 267 (1977).
%%CITATION = PHRVA,D15,267;%%
\bibitem{Lohse}
D.~Lohse~\textit{et al.},~Nucl. Phys. \textbf{A516}, 513 (1990).
%%CITATION = NUPHA,A516,513;%%
\bibitem{Weinstein}
J.~D.~Weinstein,~N.~Isgur,~Phys. Rev.~D~\textbf{41}, 2236 (1990).
%%CITATION = PHRVA,D41,2236;%%
\bibitem{Beveren}
E.~Van Beveren,~\textit{et al.}, Z. Phys.~C~\textbf{30}, 615 (1986).
%%CITATION = ZEPYA,C30,615;%%
\bibitem{Johnson}
R.~L.~Jaffe,~K.~Johnson,~Phys. Lett. \textbf{B60}, 201 (1976).
%%CITATION = PHLTA,B60,201;%%
\bibitem{Kaiser}
N.~Kaiser,~P.~B.~Siegel,~W.~Weise,~Nucl. Phys. \textbf{A594}, 325 (1995).
%%CITATION = NUPHA,A594,325;%%
\bibitem{Li}
G.~Q.~Li,~C.-H.~Lee,~G.~E.~Brown,~Nucl. Phys. \textbf{A625}, 372 (1997).
%%CITATION = NUPHA,A625,372;%%
\bibitem{Senger}
P.~Senger~\textit{et al.},~Phys. Rev.~C \textbf{75}, 024906 (2007).
%%CITATION = PHRVA,C75,024906;%%
\bibitem{Laue}
F.~Laue~\textit{et al.},~Phys. Rev. Lett. \textbf{82}, 1640 (1999).
%%CITATION = PRLTA,82,1640;%%
\bibitem{Barth}
R.~Barth~\textit{et al.},~Phys. Rev. Lett. \textbf{78}, 4007 (1997).
%%CITATION = PRLTA,78,4007;%%
\bibitem{Menzel}
M.~Menzel~\textit{et al.},~Phys. Lett. \textbf{B495}, 26 (2000).
%%CITATION = PHLTA,B495,26;%%
\bibitem{cosy}
D. Prasuhn~\textit{et al.}, Nucl. Instrum. Methods Phys. Res.~A \textbf{441}, 167 (2000);
%%CITATION = NUIMA,A441,167;%%
R. Maier,~\textit{ibid.} \textbf{390}, 1 (1997).
%%CITATION = NUIMA,A390,1;%%
\bibitem{review}
P.~Moskal, M.~Wolke, A.~Khoukaz, W.~Oelert, Prog. Part. Nucl. Phys. \textbf{49}, 1 (2002).
%%CITATION = PPNPD,49,1;%%
\bibitem{wolke}
M.~Wolke,~PhD thesis, IKP J{\"u}l-3532 (1997).
%% CITATION = IKP J\"ul-3532;%%
\bibitem{quentmeier}
C.~Quentmeier~\textit{et al.},~Phys. Lett. \textbf{B515}, 276 (2001).
%%CITATION = PHLTA,B515,276;%%
\bibitem{winter}
P.~Winter~\textit{et al.},~Phys. Lett. \textbf{B635}, 23 (2006).
%%CITATION = PHLTA,B635,23;%%
\bibitem{anke}
Y.~Maeda~\textit{et al.},~Phys. Rev.~C \textbf{77}, 01524 (2008).
%%CITATION = PHRVA,C77,015204;%%
\bibitem{c_wilkin}
C.~Wilkin,~AIP Conf. Proc. \textbf{950}, 23 (2007).
%%CITATION = APCPC,950,23;%%
\bibitem{disto}
F.~Balestra~\textit{et al.},~Phys. Lett. \textbf{B468}, 7 (1999).
%%CITATION = PHLTA,B468,7;%%
\bibitem{goldhaber1}
W.~Chinowsky, G.~Goldhaber, S.~Goldhaber, W.~Lee, T.~O'Halloran,
Phys. Rev. Lett. \textbf{9}, 330 (1962).
%%CITATION = PRLTA,9,330;%% 
\bibitem{goldhaber2}
W.~Chinowsky, G.~Goldhaber, S.~Goldhaber, W.~Lee, T.~O'Halloran, Phys. Rev. Lett. \textbf{6}, 62 (1963).
%%CITATION = PRLTA,6,62;%%
\bibitem{wilkin2007}
C. Wilkin, remarks at the Symposium on Meson Physics, Cracow (2007).
\bibitem{oelert1} 
W.~Oelert~\textit{et al.},~Int. J. of Mod. Phys.~A \textbf{22}, 502 (2007).
%%CITATION = IMPAE,A22,502;%%
\bibitem{pp-FSI}
P.~Moskal~\textit{et al.},~Phys. Lett. \textbf{B482}, 356 (2000).
%%CITATION = PHLTA,B482,356;%%
\bibitem{c-11}
S.~Brauksiepe~\textit{et al.}, Nucl. Instrum. Methods Phys. Res.~A \textbf{376}, 397 (1996).
%%CITATION = NUIMA,A376,397;%%
\bibitem{dombrowski}
 H. Dombrowski \textit{et al.}, Nucl. Instrum. Methods Phys. Res.~A~\textbf{386}, 228 (1997).
%%CITATION = NUIMA,A386,228;%%
\bibitem{moskal1}
P.~Moskal~\textit{et al.},~J. Phys. G \textbf{28}, 1777 (2002).
 %%CITATION = JPHGB,G28,1777;%% 
\bibitem{noyes995}
H.~P.~Noyes, H.~M.~Lipinski,~Phys. Rev.~C \textbf{4}, 995 (1971).
%%CITATION = PHRVA,C4,995;%%
\bibitem{noyes465}
H.~P.~Noyes,~Ann. Rev. Nucl. Sci. \textbf{22}, 465 (1972).
%%CITATION = ARNUA,22,465;%%
\bibitem{naisse506}
J.~P.~Naisse,~Nucl. Phys. \textbf {A278}, 506 (1977).
%%CITATION = NUPHA,A278,506;%%
\bibitem{habilitacja}
P.~Moskal,~\textit{e-Print Archive}: hep-ph/0408162.
%%CITATION = HEP-PH/0408162;%%
\bibitem{nyborg}
P.~Nyborg~\textit{et al.},~Phys. Rev. \textbf{140}, 914 (1965).
%%CITATION = PHRVA,140,B914;%% 
\bibitem{chodrow}
D.~Chodrow,~Nuovo Cimento \textbf{50}, 674 (1967).
\bibitem{kaminski}
R.~Kami\'nski, L. Le\'sniak,~Phys. Rev.~C \textbf{54}, 2264 (1995).
 %%CITATION = PHRVA,C54,2264;%%
\bibitem{Baru}
V.~Baru~\textit{et al.}, Phys. Lett.~\textbf{B586}, 53 (2004).
%%CITATION = PHLTA,B586,53;%%
\bibitem{Teige}
S.~Teige~\textit{et al.}, Phys. Rev.~D \textbf{59}, 012001 (2001).
%%CITATION = PHRVA,D59,012001;%%
\bibitem{Bugg} 
D. V.~Bugg, V. V.~Anisovich, A.~Sarantsev, B.S.~Zou, Phys. Rev.~D \textbf{50}, 4412 (1994).
%%CITATION = PHRVA,D50,4412;%%
\bibitem{mich_mgr}
M.~Silarski, arXiv:0809.0837v1; FZ-J\"ulich report, J\"ul-4278, (2008).%, [arXiv:0809.0837]
%%CITATION = IKP J\"ul-4278
%%CITATION = ARXIV:0809.0837;%%
\bibitem{bakernim}
S. Baker, R.D. Cousins,~Nucl. Instrum. Methods Phys. Res.~A \textbf{221}, 437 (1984).
%%CITATION = NUIMA,221,437;%%
\bibitem{feldmanpr}
G.~J. Feldman, R.~D. Cousins, Phys. Rev.~D \textbf{57}, 3873 (1998).
%%CITATION = PHRVA,D57,3873;%%
\bibitem{dzyuba}
A. Dzyuba~\textit{et al.}, Phys. Lett.~\textbf{B668}, 315 (2008).
%%CITATION = PHLTA,B668,315;%%
\end{thebibliography}
\end{document}